\documentclass[onefignum,onetabnum]{siamart171218}

\usepackage{epstopdf}
\ifpdf
  \DeclareGraphicsExtensions{.eps,.pdf,.png,.jpg}
\else
  \DeclareGraphicsExtensions{.eps}
\fi

\newsiamremark{remark}{Remark}
\newsiamremark{hypothesis}{Hypothesis}
\crefname{hypothesis}{Hypothesis}{Hypotheses}
\newsiamthm{claim}{Claim}

\usepackage{setspace}

\usepackage{amsmath,amssymb,amsfonts,mathrsfs}

\usepackage[english]{babel}

\usepackage{hyperref}

\usepackage{graphicx}
\usepackage{color}
\usepackage{pdfpages}

\graphicspath{{./Figures/}} 
\usepackage[hangindent=0pt,singlelinecheck=on,format=plain,font=scriptsize]{caption}
\usepackage[font=scriptsize,labelformat=parens]{subcaption}

\usepackage{float}
\restylefloat{figure}
\usepackage{wrapfig}

\newcommand {\real}{\mathbb{R}}

\newcommand {\tor}{\mathbb{T}}
\newcommand {\id}{\mathbb{I}}

\newcommand {\peclet}{\mathrm{Pe}}

\newcommand {\eps} {\varepsilon}
\newcommand {\vphi} {\varphi}

\DeclareMathOperator{\linhull}{span}
\DeclareMathOperator*{\minimize}{minimize}

\newcommand{\DIV}{\nabla\cdot} 

\newcommand{\T}{\mathcal{T}} 
\newcommand{\R}{\mathcal{R}} 

\newcommand{\set}[1]{\left\lbrace #1 \right\rbrace}
\newcommand{\ve}[1]{\mathbf{#1}}

\newcommand{\norm}[2]{\left\| #1 \right\|_{#2}}

\renewcommand{\d}{\:\mathrm{d}}

\newcommand{\E}[1]{\cdot 10^{#1}}

\usepackage{algorithm} 
\usepackage{algorithmic}  
\usepackage[algo2e,algoruled,linesnumbered]{algorithm2e}
\usepackage{setspace}
\makeatletter
\newcommand{\removelatexerror}{\let\@latex@error\@gobble}
\makeatother

\headers{Semi-Lagrangian Multiscale Reconstruction}{K. Simon, and J. Behrens}

\title{Semi-Lagrangian Subgrid Reconstruction for Advection-Dominant
  Multiscale Problems\thanks{Submitted \today.  \funding{This research
      was conducted in the framwork of the PalMod project funded by
      the German Ministry of Education and Research (BMBF).}}}

\author{Konrad Simon\thanks{University of Hamburg, Department of
    Mathematics, Grindelberg 5, 20144 Hamburg, Germany
    (\email{konrad.simon@uni-hamburg.de}).}  \and J\"orn
  Behrens\thanks{University of Hamburg, Department of Mathematics,
    Grindelberg 5, 20144 Hamburg, Germany
    (\email{joern.behrnes@uni-hamburg.de}).}}

\begin{document}

\maketitle

\begin{abstract}
  We introduce a new framework of numerical multiscale methods for
  advection-dominated problems motivated by climate sciences. Current
  numerical multiscale methods (MsFEM) work well on stationary
  elliptic problems but have difficulties when the model involves
  dominant lower order terms. Our idea to overcome the assocociated
  difficulties is a semi-Lagrangian based reconstruction of subgrid
  variablity into a multiscale basis by solving many local inverse
  problems. Globally the method looks like a Eulerian method with
  multiscale stabilized basis. We show example runs in one and two
  dimensions and a comparison to standard methods to support our ideas
  and discuss possible extensions to other types of Galerkin methods,
  higher dimensions and nonlinear problems.
\end{abstract}



\section{Introduction}\label{s-1}

\subsection{Motivation and Overview}\label{s-1-1}

Simulating complex physical processes at macroscopic coarse scales
poses many problems to engineers and scientists. Such simulations
strive to reflect the effective behavior of observables involved at
large scales even if the processes are partly driven by highly
heterogenous micro scale behavior. On the one hand hand resolving the
microscopic processes would be the safest choice but such a strategy
is prohibitive since it would be computationally expensive. On the
other hand microscopic processes significantly influence the
macroscopic behavior and can not be neglected.

Incorporating micro scale effects into macro simulations in a
mathematically consistent way is a challenging task. There exist many
scenarios in different disciplines of science that are faced with such
challenges. In fully coupled paleo climate simulations, i.e., climate
simulations over more than hundred thousand years a typical grid cell
has edge lengths around 200 kilometres and more. Consequently, subgrid
processes such as heterogeneously distributed and moving ice shields
are not or just insufficiently resolved~\cite{Notz2017}. These subgrid
processes are usually taken care of by so-called
parameterizations. One can imagine this as small micro scale
simulations that are then coupled to the prognostic variables such as
wind speed, temperature and pressure on the coarse grid (scale of the
dynamical core). This coupling from fine to coarse scales is being
referred to as upscaling and is unfortunately often done in rather
heuristic ways. This leads to wrong macroscopic quantities like wrong
pressures and eventually even to wrong wind directions and more
undesired effects such as phase errors.

The increasing complexity of earth system models (ESMs), in general,
demands for mathematically consistent upscaling of parametrized
processes that occur or are modelled on very different scales relative
to the dynamical core. A first example was already mentioned: In
global climate simulations a sea ice model computes the average ice
cover for each coarse cell. This information then enters heat fluxes
between ocean and atmosphere since sea ice forms an interface between
them and is hence modeled as an averaged diffusive process. It is
known from homogenization theory that simple averaging of heterogenous
diffusive quantities does not reflect the correct coarse scale
diffusion~\cite{Bensoussan2011}.

Another, quite contrary, approach to perfom multiscale simulations is
adaptive mesh refinement (AMR), see~\cite{Behrens2006}. The idea of
AMR is to initially run a coarse scale simulation and then assess the
quality of the solution locally by means of mathematically or
physically based a posteriori estimators. The coarse mesh is then
refined in regions where the quality of the solution is not sufficient
or coarsened where the solution is smooth. This approach is different
from the above upscaling since micro scales are locally resolved -- it
is therefore sometimes referred to as downscaling. Such an approach is
attractive and is being used in practice but it is not feasible in
situations in which heterogeneities in the solution are expected to
occur globally.

One early attempt to use multiscale methods are Eulerian Lagrangian
localized adjoint methods (ELLAM), which constitute a space-time
finite element framework, see~\cite{Celia1990,Herrera1993}, and form
the basis of their multiscale version
MsELLAM~\cite{Wang2009,Cheng2010}. Such methods rely on operator
splitting and basis functions are required to satisfy an adjoint
equation. For a review see~\cite{Russell2002}.

There exist many other multiscale methods. Homogenization is an
originally analytical tool to find effective models of otherwise
heterogenous models in the limit of a large scale
separation~\cite{Allaire1992,Bensoussan2011,Jikov2012}. Unfortunately
it is often too difficult to find such an effective equation but there
exist numerical techniques that aim at designing numerical algorithms
to effectively capture the behavior of the solution to the unknown
homogenized problem.

The heterogenous multiscale method (HMM) was pioneered by E and
Enquist~\cite{Weinan2003,WeinanEnquistHuang2003} and refers to a rich
population of variants~\cite{Abdulle2007,Henning2010,Henning2014}. This method
emphasizes important principles in the design of multiscale methods
such as the choice of macro- and micro-models and their
communication. For reviews and further references
see~\cite{Abdulle2012,Weinan2011} and~\cite{Abdulle2009} for a
discussion of the HMM with special focus on PDEs.

Variational multiscale methods (VMM) have been developed in the 1990s
by Hughes and collaborators, see~\cite{Hughes1995, Hughes1998}. The
spirit of the method lies in a decomposition of the solution space and
in the design of variational forms that reflect the relevant scale
interactions between the spaces. Many versions of it exist and we
point the reader to recent reviews~\cite{Ahmed2017,Rasthofer2018} and
the vast literature therein.

The present work is inspired by multiscale finite element methods
(MsFEM) which can be seen as a special variant of the VMM. The idea
of this method is to introduce subgrid variations into basis functions
and can be dated back to works by Babu{\v{s}}ka, Caloz and
Osborn~\cite{Babuska1994} and shares ideas with the partition of unity
method~\cite{Melenk1996}. The MsFEM in its current form was introduced
in~\cite{HouWu1997,HouWu1999, Efendiev2000}. The essential idea of the
method is to capture the local asymptotic structure of the solution
through adding problem dependent bubble correctors to a standard basis
and use these as ansatz and trial functions. Many variations of this
method exist and refer the reader to~\cite{Efendiev2009, Graham2012}
for a review.

Many of the afore mentioned methods have the advantage that they work
well for elliptic or parabolic problems and that they are accessible
to an analysis. The difficulty in many applications on the other hand
is their advection or reaction dominated character, i.e., the dynamics
is often driven by low order terms. This poses major difficulties to
numerical multiscale methods. Multiscale finite element methods
naively applied will not converge to any reasonable solution since
basis functions will exhibit artificial boundary layers that are not
present in the actual physical flow. Ideas to tackle this problem are
based on combining transient multiscale methods with Lagrangian
frameworks~\cite{Simon2018} or with stabilization methods for
stationary problems, see~\cite{LeBris2017} for an overview. A HMM
based idea for incompressible turbulent flows can be found
in~\cite{Lee2016}. For a method based on the VMM see~\cite{Li2016}.

\subsection{Contribution}\label{s-1-2}

Our main contribution is a framework of numerical methods for
advection-dominated flows which by reconstructing subgrid variations
on local basis functions aims at reflecting the local asymptotic
structure of solutions correctly. The idea combines multiscale
Galerkin methods with semi-Lagrangian methods by locally solving an
inverse problem for the basis representation of solutions that is
adapted to the actual flow scenario. We demonstrate the idea on one
and two-dimensional advection-diffusion equations with heterogenous
background velocities and diffusivities in both non-conservative and
conservative form.

Applying standard MsFEMs directly leads to failure since boundary
conditions (BCs) for the modified basis functions can not be
prescribed arbitrarily. Suitable BCs on the subgrid scale are an
essential ingredient for many multiscale methods. A wrong placement of
Dirichlet BCs, for example, leads to boundary layers in basis
functions that are not there in the real large scale flow, i.e., the
way information propagates in advection-dominant flow needs to be
respected and not artificially blocked. Other choices of BCs usually
complicate the enforcement of conformity or obfuscate additional
assumptions on the problem structure (such as local periodicity).

Conformal MsFEM techniques for advection-dominated tracer transport
were already explored in our previous work in one spatial dimension on
a transient advection-diffusion equation \cite{Simon2018}. The finding is that one has
to follow a Lagrangian point of view on coarse scales such that flow
is ``invisible''.  On fine scales one can then simplify Lagrangian
transforms in order to make advective effects locally milder without
going to a fully Lagrangian setting. This amounts to prescribing
Dirichlet BCs on coarse flow characteristics and has the effect that
basis functions do not develop spurious boundary layers not present in
the actual flow. While this work gave some useful insights it is
unfortunately not feasible for practical applications since it suffers
from several weaknesses.  First, it is not directly generalizable to
higher dimensions. Secondly, it needs assumptions on the background
velocity that are not necessarily fulfilled in practical applications
to ensure that coarse scale characteristics do not merge.

In order to circumvent these problems we suggest here a new idea based
on a semi-Lagrangian framework that locally in time constructs a
multiscale basis on a fixed Eulerian grid. The construction is done in
a semi-Lagrangian fashion on the subgrid scale whereas the macroscopic
scale is conveniently treated fully Eulerian. This is in
complete contrast to our previous work but still respects that
information in advection-dominant flows is ``mostly'' propagated along
flow characteristics.

The construction of the basis in each cell at time $t^{n+1}$ is done
by tracing each Eulerian cell back to time $t^n$. This yields a
distorted cell. A basis on this distorted cell is then reconstructed
by solving an inverse problem for the representation of the global
solution at the previous time step. Then the local representation of
the solution on this cell is propagated forward in time to get a basis
on the Eulerian grid at $t^{n+1}$ instead of propagating the solution
itself.

All reconstructions of the basis in coarse cells are independent from
each other and can be performed in parallel. The global time step with
the modified basis is not critical since the coarse problem is small
and since we can use algebraic constructs to make the assembly of the
global problem efficient. Note that although we formulate the
algorithm globally in an implicit form it can well be formulated
explicitly which will allow for computational efficiency for the global step.

Our new approach has several advantages. First, we can effectively
incorporate subgrid behavior of the solution into the multiscale basis
via the solution of local inverse problems. Secondly, the method can
handle advection-dominated flows in a parallel fashion and,
furthermore, the idea works in any dimension. Numerical tests show
that it is accurate in both $L^2$ and $H^1$ since it represents
subgrid variability correctly. The method can 
handle problems that involve an additional reaction term. This
consequently includes conservation problems. Furthermore, the idea is
generic and can be used for vector valued problems and problems that
involve subgrid information coming from actual data.

\section{The Semi-Lagrangian Basis reconstruction in One and Two
  Spatial Dimensions}\label{s-2}

In this section we will outline our ideas on an advection-diffusion
equation (ADE) with periodic boundaries as a model problem. We will
outline all ideas for didactical purposed first in $d=1$ and in $d=2$
dimensions on
\begin{equation}
  \label{eq:2-1a}
  \begin{split}
    \partial_tu + \ve c_\delta\cdot\nabla u & = \nabla\cdot\left(\ve A_\eps\nabla u\right) + f \quad \text{in } \tor^d\times [0,T]\sim [0,1]^d\times [0,T] \\
    u(\ve x,0) & = u_0(\ve x) \\
  \end{split}
\end{equation}
and 
\begin{equation}
  \label{eq:2-1b}
    \begin{split}
      \partial_tu + \nabla\cdot \left( \ve c_\delta u\right) & = \nabla\cdot\left(\ve A_\eps\nabla u\right) + f \quad \text{in } \tor^d\times [0,T] \\
      u(\ve x,0) & = u_0(\ve x) \\
  \end{split}
\end{equation}
where $\ve c_\delta(\ve x,t)$ is the background velocity,
$\ve A_\eps(\ve x,t)$ is a matrix-valued diffusivity, and $f$ and
$u_0$ are some smooth external forcing and initial condition. We will
use bold letters for the vectors and tensors independent of the
dimension.

The indices $\delta>0$ and $\eps>0$ indicate that both quantities may
have large variations on small scales that are not resolved on coarse
scales $H>0$ of our multiscale method. We will also work locally on a
scale $h\ll H$ that can resolve the variations in the
coefficients. Furthermore, we assume that $\ve c_\delta\gg\ve A_\eps$
(see remark below) and that $\ve c_\delta$ is well-behaved, for
example $C^{1-}$ in space and continuous in time. This assumption is
often satisfied in practice for example in climate simulations where
$\ve c_\delta$ is given at the nodes of a coarse grid. Depending on
the application variations in $\ve c_\delta$ may be resolved on scale
$H$ or not. The diffusivity tensor is assumed to be positive definite
(uniformly in $\eps$ and point-wise in $\ve x$) with
$\ve A_\eps\in L_t^\infty (L_x^\infty)^{d\times d}$, and is often
derived from parametrized processes such as a varying sea ice
distribution, convection, topographical features, or land use
patterns. Note, that~(\ref{eq:2-1a}) does not conserve the tracer $u$
in contrast to equation~(\ref{eq:2-1b}) provided $f=0$.

\begin{remark}
  Advection-dominance of a flow (what we sloppily expressed by
  $\ve c_\delta\gg\ve A_\eps$) is usually expressed by a dimensionless
  number -- the P\'{e}clet number $\peclet$ which is essentially the
  ratio between advective and diffusive time scales. There exist
  several versions of this number~\cite{Huysmans2005}. Since for large
  variations of the coefficients on the subgrid scale
  advection-dominance is a very local property we need be more precise
  with what me mean by that. Here we assume that $\peclet$ is high on
  average, i.e.,
  $\peclet=\frac{\norm{\ve c_\delta}{L^2}L}{\norm{\ve A_\eps}{L^2}}$
  where $L$ is a characteristic length. We take $L$ to be the length
  of the computational domain.
\end{remark}

Although the general idea of reconstructing subgrid variability in
basis functions is not substantially different for~(\ref{eq:2-1a})
and~(\ref{eq:2-1b}) there are some differences in the propagation of
local boundary information described below that strongly influence the
accuracy. We will explicitly describe these differences.

\begin{figure}[t!]
  \begin{center}
    \includegraphics[width=1.0\textwidth]{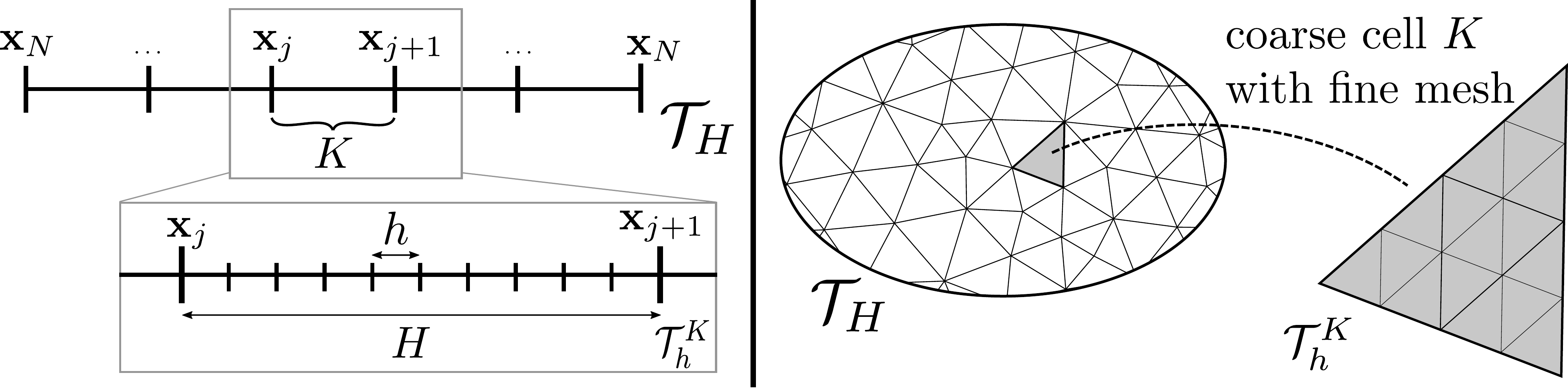}
  \end{center}
  \vspace{5pt}
  \caption{Global coarse mesh and local fine mesh in 1D (left) and 2D
    (right).}
  \label{fig-1d_multiscale_mesh}
\end{figure}

We start with outlining our method in one dimension since the idea is
very simple and avoids complications that arise in higher
dimensions. The idea is to represent non-resolved fine scale
variations of the solution locally on a set of non-polynomial basis
functions in each coarse cell. That is we fix two (Eulerian) meshes: A
coarse mesh $\T_H$ of width $H$ and on each cell $K\in\T_H$ of the
coarse mesh we have a fine mesh $\T_h^K$ of width $h\ll H$ (we assume
that $h$ does not depend on $K$). On the coarse mesh we have a
multiscale basis
$\vphi_i^{H,\mathrm{ms}}:\T^d\times [0,T]\rightarrow\real$,
$i=1,\dots,N_H$, where $N_H$ is the number of nodes of $\T_H$. This
basis depends on space and time and will be constructed so that we
will obtain a spatially $H^1$-conformal (multiscale) finite element
space. Note that this is a suitable space for problems~(\ref{eq:2-1a})
and~(\ref{eq:2-1b}) since they have unique solutions
$u\in L^2([0,T], H^1(\tor^d))$ with
$\partial_t u\in L^2([0,T], H^{-1}(\tor^d))$ and hence in
$u\in C([0,T], L^2(\tor^d))$~\cite{Evans10}. The initial condition
$\ve{u}^0$ can therefore be assumed to be in $L^2(\tor^d)$ but in
later experiments we will choose it to be smooth. The fine mesh on
each cell $K\in\T_H$ is used to represent the basis locally, see
Figure~\ref{fig-1d_multiscale_mesh}.

Standard MsFEM methods used for porous media flows are designed in
such a way that the coarse scale basis solves the PDE model locally
with the same boundary conditions as standard FEM basis
functions. Note that the global coarse scale MsFEM solution is a
linear combination of modified local basis functions that do resolve
the fine scale structure induced by the problem's heterogenities and
therefore do represent the asymptotics correctly. This works for
stationary elliptic problems and for parabolic problems as long as
there is no advective term involved or even dominant. The reason is
that advective terms as they appear in our model problems prevent a
basis constructed by a standard MsFEM technique to represent the
correct asymptotics. This is since flow of information is artificially
blocked at coarse cell boundaries. Due to the incorrect boundary
conditions for the multiscale basis artificial steep boundary layers
form in basis functions that do not exist in the actual global
flow. For transient problems another difficulty ist that the local
asymptotics around a point depends on the entire domain of dependence
of this point and hence subscale information represented by a base
function must contain memory of the entire history of the local domain
of that base function.

A first attempt to bypass these difficulties was to pose boundary
conditions for the basis on suitable space-time curves, across which
naturally no information is propagated on the coarse scales. Such
curves are Lagrangian paths. Therefore, a Lagrangian framework on the
coarse scale and an ``almost Lagranian'' setting on fine scales was
proposed by us earlier~\cite{Simon2018}. Unfortunately, this method is
not feasible since it can not be generalized to higher dimensions and
needs restrictive assumptions on the flow field.

Nonetheless, the results of~\cite{Simon2018} suggest that a coarse
numerical splitting of the domain should correspond to a resonable
physical splitting of the problem. Instead of a fully Lagrangian
method on coarse scales semi-Lagrangian techniques to build the basis
can circumvent the difficulties of Lagrangian techniques. But these
are only local in time and therefore they do not take into account the
entire domain of dependence of a point. We show how to deal with this
in the following. First, we start with the global problem.

\subsection{The global time step in 1D/2D}\label{s-2-1}

Suppose we know a set of multiscale basis functions in a conformal
finite element setting, i.e., we approximate the global solution at
each time step in a spatially coarse subspace
$V^H(t) \subset H^1(\tor^d)$ in which the solution $u$ is sought
(almost everywhere). We denote this finite-dimensional subspace as
\begin{equation}
  \label{eq:2-7}
  V^H(t) = \linhull \set{ \left. \vphi_j^{H,\mathrm{ms}}(\cdot, t) \: \right| \: i=1,\dots ,N_H \: \text{}} \:.
\end{equation}
First, we expand the solution $u^H(\ve x,t)$ in terms of the
 basis at time $t\in [0,T]$, i.e.,
\begin{equation}
  \label{eq:2-8}
  u^H(\ve x,t) = \sum_{j=0}^{N_H} u_j^H(t)\vphi_j^{H,\mathrm{ms}}(\ve x, t) \:.
\end{equation}
Then we test with the modified basis and integrate by
parts. Therefore, the spatially discrete version of both
problem~(\ref{eq:2-1a}) and~(\ref{eq:2-1b}) becomes the ODE
\begin{equation}
  \label{eq:2-9}
  \begin{split}
    \ve{M}(t)\frac{\d}{\d t}\ve{u}^{H}(t) + \ve{N}(t)\ve{u}^{H}(t) &= \ve{A}(t)\ve{u}^{H}(t) + \ve f^{H}(t) \\
    \ve{u}^H(0) &= \ve{u}^{H,0} 
  \end{split}
\end{equation}
where
\begin{multline}
  \label{eq:2-10a}
  \ve{A}_{ij}(t) = \int_{\tor^d} \vphi_i^{H,\mathrm{ms}}(\ve x, t) \cdot \ve{A}_\eps(\ve x, t)\nabla \vphi_j^{H,\mathrm{ms}}(\ve x, t) \d \ve x\\
  - \int_{\tor^d}  \vphi_i^{H,\mathrm{ms}}(\ve x, t)\left(\ve{c}_\delta(\ve x, t)\cdot\nabla \right)\vphi_j^{H,\mathrm{ms}}(\ve x, t) \d \ve x 
\end{multline}
for~(\ref{eq:2-1a}) and
\begin{multline}
  \label{eq:2-10b}  
  \ve{A}_{ij}(t) = \int_{\tor^d} \vphi_i^{H,\mathrm{ms}}(\ve x, t) \cdot \ve{A}_\eps(\ve x, t)\nabla \vphi_j^{H,\mathrm{ms}}(\ve x, t) \d \ve x\\
  + \int_{\tor^d}  \left(\nabla\vphi_i^{H,\mathrm{ms}}(\ve x, t)\cdot\ve{c}_\delta(\ve x, t)\right) \vphi_j^{H,\mathrm{ms}}(\ve x, t) \d \ve x
\end{multline}
for~(\ref{eq:2-1b}). The mass matrix is given by
\begin{equation}
  \label{eq:2-11}  
  \ve{M}_{ij}(t) = \int_{\tor ^d}\vphi_i^{H,\mathrm{ms}}(\ve x, t)\vphi_j^{H,\mathrm{ms}}(\ve x, t) \d \ve x \:,
\end{equation}
$\ve{f}^{H}(t)$ contains forcing and boundary conditions and the
initial condition $\ve{u}^{H,0}$ is the projection of
$\ve{u}^0\in L^2(\tor^d)$ onto $V^H(0)$.  Note that~(\ref{eq:2-9})
contains a derivative of the mass matrix:
\begin{equation}
  \label{eq:2-12}
  \ve{N}_{ij}(t) = \int_{\tor ^d}\vphi_i^{H,\mathrm{ms}}(\ve x, t) \partial_t\vphi_j^{H,\mathrm{ms}}(\ve x, t) \d \ve x \:.
\end{equation}
This is necessary since the basis functions depend on time and since
we discretized in space first. The reader will notice that using
Rothe's method of lines, i.e., discretizing in time first, it is not
clear what basis function to use for testing and this a priori leads
to a different linear system.

\begin{figure}[t!]
  \vspace{0pt}
  \begin{center}
    \includegraphics[width=0.99\textwidth]{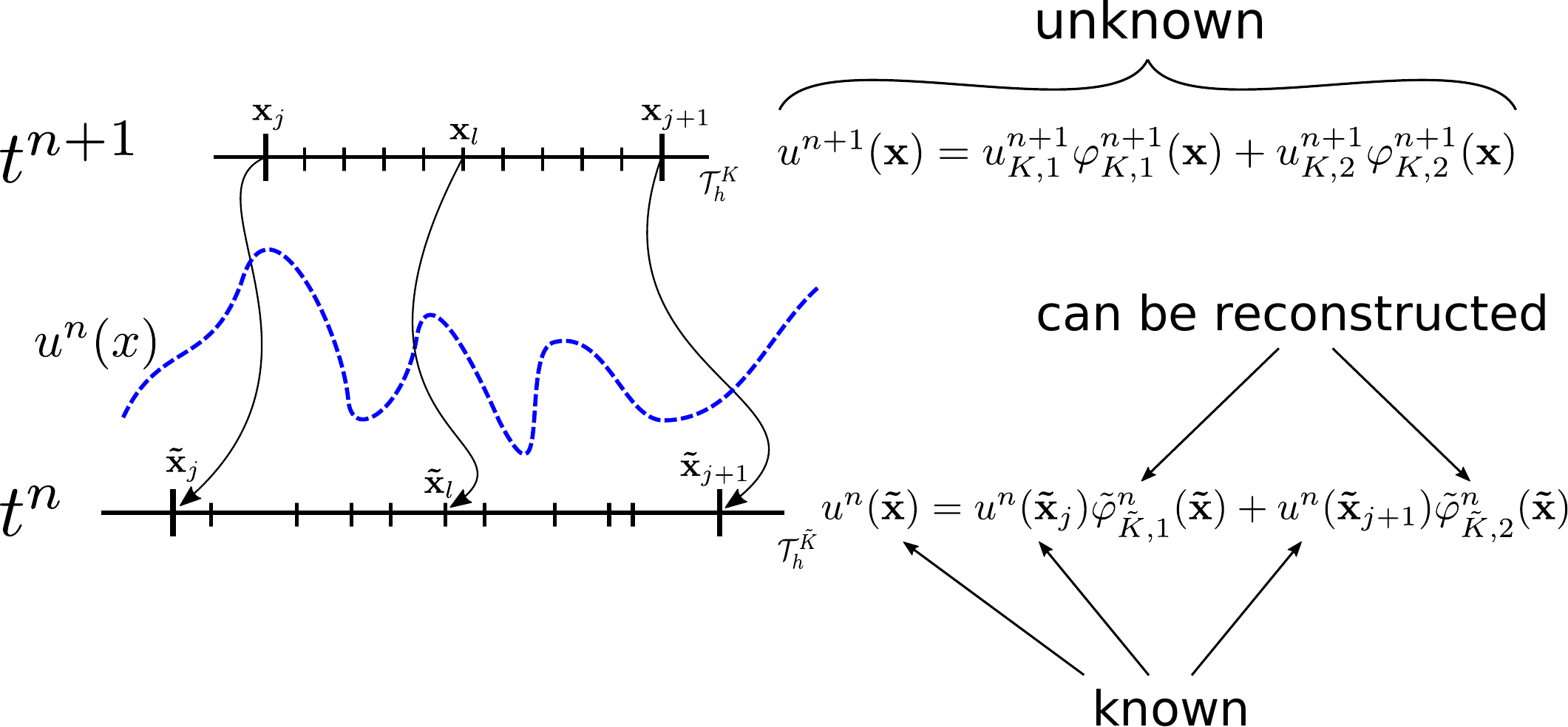}
  \end{center}
  \vspace{0pt}
  \caption{The fine mesh in each cell $K\in\T_H$ is traced back one
    time step where the known solution can be used to reconstruct a
    basis representation of the solution.}
  \vspace{0pt}
  \label{fig-1d_traceback}
\end{figure}

For the time discretization we simply use the implicit Euler
method. The discrete ODE then reads
\begin{equation}
  \label{eq:2-13}
  \ve{M}(t^{n})\ve{u}^{n+1} 
  = \ve{M}(t^{n})\ve{u}^{n} + \delta t \left[ \ve{A}(t^{n+1})\ve{u}^{n+1} - \ve{N}(t^{n+1})\ve{u}^{n+1} + \ve f^{H}(t^{n}) \right] \:.
\end{equation}
Other time discretization schemes, in particular, explicit schemes are
of course possible. For didactic reasons we choose to present the
algorithm in an implicit version. We will come back to that later. The
next step is to show how to construct the multiscale
\begin{wrapfigure}[20]{r}{0.59\textwidth}
  \vspace{0pt}
  \begin{center}
    \includegraphics[width=0.59\textwidth]{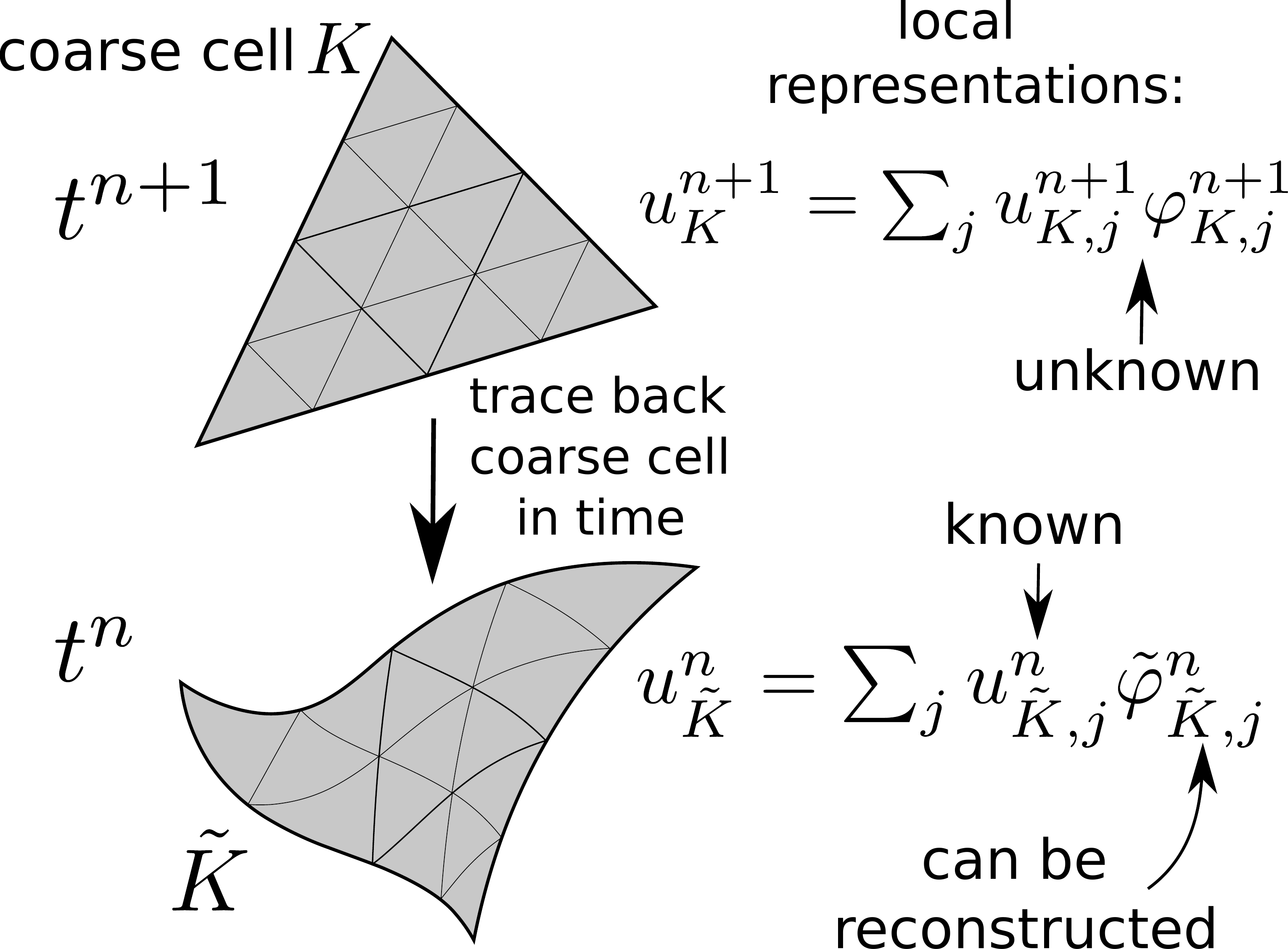}
  \end{center}
  \vspace{0pt}
  \caption{2D illustration similar to the one shown in
    Figure~\ref{fig-1d_traceback}. Each coarse cell is together with
    its fine meshbeing traced back one time step at which the known
    global solution can be used to reconstruct a basis.}
  \vspace{0pt}
  \label{fig-2d_traceback}
\end{wrapfigure} basis.

\textbf{Convention.} In both 1D and 2D we use bold letters like
$\ve{x}, \ve{A}$ to express potentially vector valued quantities
although they would be scalar in one dimension in all formulas and
figures. We do so since we would like to avoid confusing the reader
with different notations that are merely due to different
dimensionality although the basic ideas are the same. Difficulties
that arise in more than one dimension will be pointed out
explicitly. For the sake of consistency we also use $\nabla$ instead
of $\partial_x$ in one dimension. Quantities marked with a tilde like
$\ve{\tilde x}$ signalize (semi-)Lagrangian quantities.

\subsection{The Reconstruction Mesh in 1D/2D}\label{s-2-2}

Our idea combines the advantage of both semi-Lagrangian and multiscale
methods to account for dominant advection. The reconstruction method
is based on the simple observation that local information of the
entire domain of dependence is still contained in the global solution
at the previous timestep. This can be used to construct an Eulerian
multiscale basis: we trace back an Eulerian cell $K\in\T_H$ at time
$t^{n+1}$ on which the solution and the basis are unknown to the
previous time step $t^n$. This gives a distorted cell $\tilde K$ over
which the solution $u^n$ is known but not the multiscale basis
$\tilde\vphi_i$, $i=1,2$.

In order to find the points where transported information originates
we trace back all nodes in $\T_H^K$ from time $t^{n+1}$ to
$t^{n}$. For this one simply needs to solve an ODE with the
time-reversed velocity field that reads
\begin{equation}
  \label{eq:2-2}
  \begin{split}
    \frac{\d}{\d t}\ve{\tilde x}_l(t) & = -\ve c_\delta (\ve{\tilde x}_l(t),-t) \:, \quad t\in [-t^{n+1}, -t^n] \\
    \ve{\tilde x}_l(-t^{n+1}) & = \ve{x}_l
  \end{split}
\end{equation}
for each $\ve x_l$ and then take
$\ve{\tilde x}_l=\ve{\tilde x}_l(-t^n)$, see
Figures~\ref{fig-1d_traceback} and~\ref{fig-2d_traceback} for an
illustration. This procedure is standard in semi-Lagrangian schemes
and can be parallelized.

\subsection{Basis Reconstruction in 1D} \label{s-2-3}

After tracing back each point $\ve{x}_l$ of $K\in\T_H$ to its origin
$\ve{\tilde x}_l$ in a distorted coarse cell $\tilde K\in\T_H$ we need
to reconstruct a local representation of the (known) solution $u^n$ on
$\tilde K$:
\begin{equation}
  \label{eq:2-3}
  \left. u^n(\ve x)\right|_{\tilde K} = u^n(\ve{\tilde x}_j)\tilde \vphi_{K,1}(\ve x) + u^n(\ve{\tilde x}_{j+1})\tilde \vphi_{K,2}(\ve x)
\end{equation}
\begin{wrapfigure}[17]{l}{0.6\textwidth}
  \vspace{0pt}
  \begin{center}
    \includegraphics[width=0.59\textwidth]{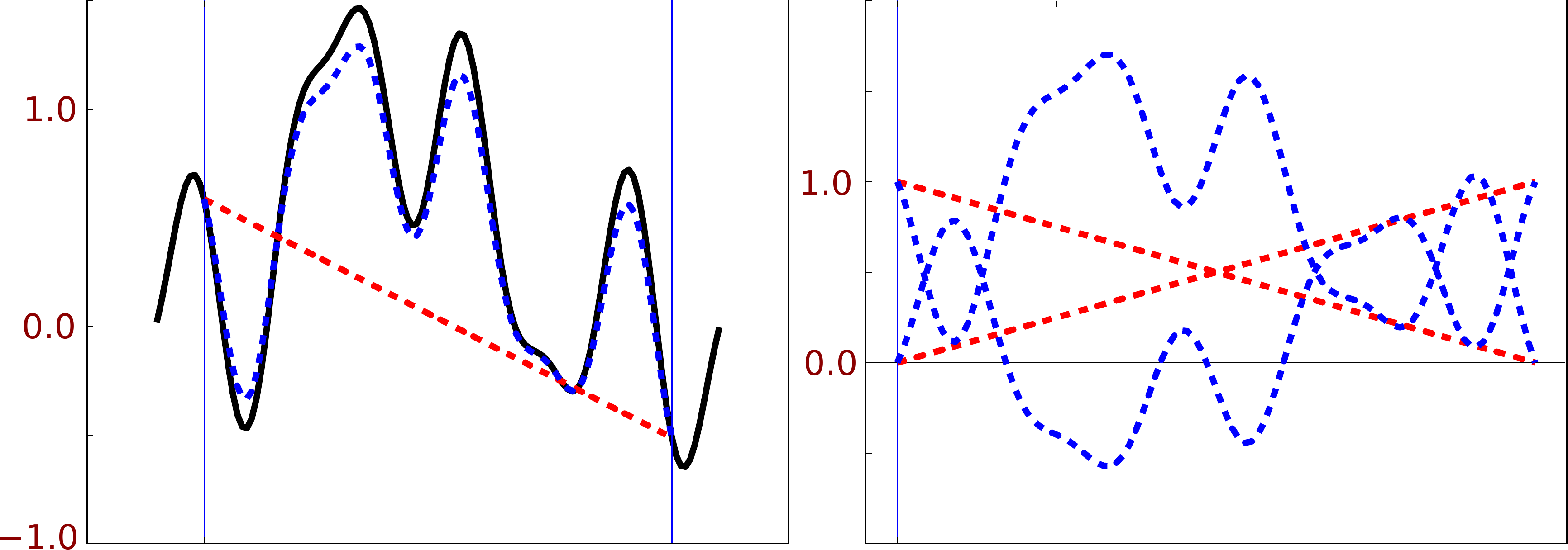}
  \end{center}
  \vspace{0pt}
  \caption{Left: Illustration of an oscillatory function (black) being
    approximated by a standard linear basis (red) on a single cell
    compared to a modified basis (blue) that
    solves~(\ref{eq:2-4}). The regularization parameters were taken as
    $\alpha_i=0.1$. Right: Comparison of the standard basis to the
    modified basis. The modified basis does neither constitute a
    partition of unity not is it positive.}
  \vspace{5pt}
  \label{fig-1d_basis_reconstruct}
\end{wrapfigure} where $\ve{\tilde x}_j$ and
$\ve{\tilde x}_{j+1}$ are the boundary points of $\tilde K$. In one
dimension one can of course choose a representation using the standard
basis of hat functions but this would not incorporate subgrid
information at step $t_n$ at all. Even in the forward advection step
(explained below) that is being carried out in the next step subgrid
information on the basis will not contain the correct subgrid
variability because the information contained in the basis does not
take into account the entire domain of dependence of $K$. We choose to
bypass this problem by solving an inverse problem for the basis to
adjusts the representation. The idea is to fit a linear combination of
the basis locally such that $u^n$ is optimally represented, i.e., we
solve
\begin{equation}
  \label{eq:2-4}
  \begin{split}
    \minimize_{\tilde\vphi_{K,i}\in C^0(\tilde K)} & 
    \quad  \norm{u^n - \left(u_j^n \tilde \vphi_{\tilde K ,1} + u_{j+1}^n \tilde \vphi_{\tilde K ,2}\right)}{L^2(\tilde K)}^2 + \sum_i\alpha_i\R_i(\tilde\vphi_{\tilde K,i})\\
    \text{s.t.} & 
    \quad u_j^n = u^n(\ve{\tilde x}_{j}) \:, \quad u_{j+1}^n = u^n(\ve{\tilde x}_{j+1}) \\
    & \quad \vphi_{\tilde K ,1}(\ve{\tilde x}_{j}) = \vphi_{\tilde K ,2}(\ve{\tilde x}_{j+1}) = 1 \\
    & \quad \vphi_{\tilde K ,1}(\ve{\tilde x}_{j+1}) = \vphi_{\tilde K ,2}(\ve{\tilde x}_{j}) = 0 \:.\\
  \end{split}
\end{equation}
The functions $\mathcal{R}_i : C^0(\tilde K)\rightarrow \real$ are
regularizers weighted by positive numbers $\alpha_i\in\real$. A simple
regularizer that we found useful in one spatial dimension is a
penalization of the deviation of the modified basis function from the
standard linear basis function with the same boundary values in the
quadratic mean, i.e., we use
\begin{equation}
  \label{eq:2-5}
  \R_i(\tilde\vphi_{\tilde K,i}) = \norm{\tilde\vphi_{\tilde K,i} - \tilde\vphi_{\tilde K,i}^0}{L^2}^2
\end{equation}

where $\tilde\vphi_{\tilde K,i}$ denotes the $t$-th standard (linear)
basis on $\tilde K$. This amounts to a system of linear equations that
can be computed explicitly. In a spatially discrete version this
system will be small and cheap to solve. A suitable choice of a
regularizer \begin{wrapfigure}[19]{r}{0.6\textwidth}
  \vspace{0pt}
  \begin{center}
    \includegraphics[width=0.6\textwidth]{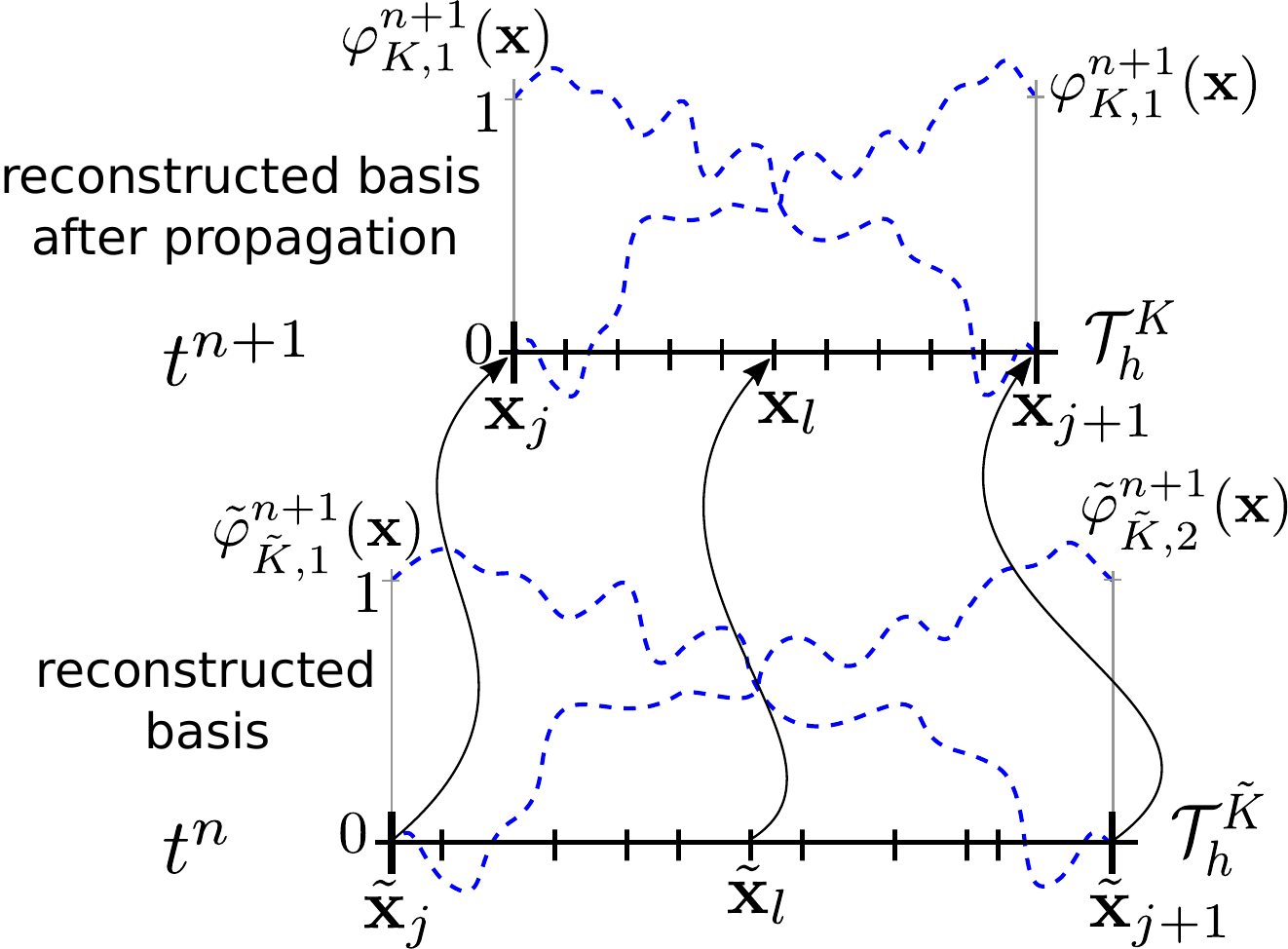}
  \end{center}
  \vspace{0pt}
  \caption{The basis reconstructed according to~(\ref{eq:2-4}) at time
    $t^n$ is propagated forward to time $t^{n+1}$ according
    to~(\ref{eq:2-6a}) or~~(\ref{eq:2-6b}).}
  \vspace{0pt}
  \label{fig-1d_propagate}
\end{wrapfigure} depends on the problem at
hand. For problems in two dimensions we choose a different one (see
below). Figure~\ref{fig-1d_basis_reconstruct} illustrates the effect
of a local reconstruction of a basis compared to a representation with
a standard basis.

\subsection{Basis Propagation in 1D}\label{s-2-4}

After having reconstructed a suitable basis on each coarse cell
$\tilde K$ we have an $H^1$-conformal basis. This basis, however is a
basis at time step $t^n$ and does not live on the coarse Eulerian grid
$\T_H$ that we initially fixed. The step to take now is to construct a
basis at $t^{n+1}$ on $\T_H$. This is done similarly
to~\cite{Simon2018}, i.e., we evolve the basis according to the model
at hand with a vanishing external forcing. Note, however, that we
compute the basis at $t^{n+1}$ along Lagrangian trajectories starting
from $t^n$, i.e., we need to transform the original
model. Equation~(\ref{eq:2-1a}) becomes
\begin{equation}
  \label{eq:2-6a}
  \begin{split}
    \frac{\d}{\d t}\vphi_{K,i} 
    & = \tilde\nabla\cdot\left( \ve{\tilde A}_\eps\tilde\nabla \vphi_{K,i} \right) \quad \text{in } \tilde K\times [ t^n,t^{n+1} ] \\
    \vphi_{K,i}(\ve{\tilde x}_j,t) 
    & = \tilde\vphi_{\tilde K,i}(\ve{\tilde x}_j) \:, \quad t\in [ t^n,t^{n+1} ] \\
    \vphi_{K,i}(\ve{\tilde x}_{j+1},t) 
    & = \tilde\vphi_{\tilde K,i}(\ve{\tilde x}_{j+1}) \:, \quad t\in [ t^n,t^{n+1} ] \\
    \vphi_{K,i}(\ve{\tilde x},t^n) 
    & = \tilde\vphi_{\tilde K,i}(\ve{\tilde x})
  \end{split}
\end{equation}
transforms into
\begin{equation}
  \label{eq:2-6b}
  \begin{split}
    \frac{\d}{\d t}\vphi_{K,i} + \left( \tilde\nabla \cdot\ve{\tilde c}_\delta\right) \vphi_{K,i}
    & = \tilde\nabla\cdot\left(\ve{\tilde A}_\eps\tilde\nabla \vphi_{K,i}\right) \quad \text{in } \tilde K\times [t^n,t^{n+1}] \\
    \vphi_{K,i}(\ve{\tilde x}_j,t) 
    & = \tilde\vphi_{\tilde K,i}(\ve{\tilde x}_j) \:, \quad t\in [ t^n,t^{n+1} ] \\
    \vphi_{K,i}(\ve{\tilde x}_{j+1},t) 
    & = \tilde\vphi_{\tilde K,i}(\ve{\tilde x}_{j+1}) \:, \quad t\in [ t^n,t^{n+1} ] \\
    \vphi_{K,i}(\ve{\tilde x},t^n) 
    & = \tilde\vphi_{\tilde K,i}(\ve{\tilde x})  \:.
  \end{split}
\end{equation}
Note that these evolution equations are solved on $\tilde K$, i.e., on
the element $K\in\T_H$ traced back in time. Advection is ``invisible''
in these coordinates. The end state
$\vphi_{K,i}(\ve{\tilde x},t^{n+1})$ on $\tilde K$ can then be
transformed onto the Eulerian element $K\in\T_H$ to obtain the desired
basis function
$\vphi_{K,i}(\ve{x},t^{n+1})\sim \vphi_{K,i}^{n+1}(\ve{x})$ at the
next time step. Corresponding basis functions in neighboring cells can
then be glued together to obtain a modified global basis
$\vphi_i^{H,\mathrm{ms}}$, $i=1,\dots,N_H$. This way we get a basis of
a subspace of $H^1$ that is neither a partition of unity nor is it
necessarily positive. Nonetheless, it is adjusted to the problem and
the data at hand. The propagation step is illustrated in
Figure~\ref{fig-1d_propagate}.

\removelatexerror
\begin{algorithm}[t!]
  \setstretch{1}
  \RestyleAlgo{ruled}
  \SetAlgoLined
  \LinesNumbered
  \SetKwInOut{Input}{input}
  \SetKwInOut{Output}{output}
  \SetKwFor{ForP}{for}{do in parallel}{end}
  
  \Input{Problem parameters for~(\ref{eq:2-1a}) or~(\ref{eq:2-1b}),
    $H$, $h$, $\delta t$}
  
  \Output{Weights $\ve u^H(t_j)$ of multiscale solution and set of
    multiscale basis functions $\vphi_i^{H,\mathrm{ms}}$}
  
  \Begin{    
    Initialize a coarse mesh $\T_H$\;
    
    On each cell $K\in\T_H$ initialize a fine mesh $\T_h^K$\;
    
    \tcc{We need to reconstruct a basis at time $t^0$. Note that no 
      trace back and no propagation are needed.}

    \ForP{$K\in\T_H$}{
      
      Reconstruct the optimal basis representation of
      $\left. u_0(x)\right|_K$ according to~(\ref{eq:2-4})\;

    }

    \tcc{Now the time stepping starts. We compute the solution at $t^{n+1}$.}
    \For{$n=0$ \KwTo $n\leq N_{\mathrm{steps}}$}{
      
      \ForP{$K\in\T_H$}{ 
        
        Trace back each node in $K$ one time step from $t^{n+1}$ to
        $t^n$ according to~(\ref{eq:2-2})\;
        
        Reconstruct the optimal basis representation of
        $\left. u^n(x)\right|_{\tilde K}$ according
        to~(\ref{eq:2-4})\;

        Propagate the optimal basis forward onto $K$ according
        to~(\ref{eq:2-6a}) or~(\ref{eq:2-6b})\; }

      Assemble the global (coarse) system matrices for~(\ref{eq:2-13}) \;

      Make a global backward Euler time step using~(\ref{eq:2-13})\;

    }

    Postprocess the solution\;

    Return\;

  }
  
  \caption{Sketch of reconstruction algorithm for a 1D
    problem. Implicit version using backward Euler. Note that all
    loops over cells $K\in\T_H$ can be parallelized.}
  \label{algo:oneD}
\end{algorithm}

Using our method we reconstruct and advect the representation of the
global solution first and then the solution itself using the modified
representation. The global step is completely Eulerian while the local
reconstruction step is semi-Lagrangian in contrast to~\cite{Simon2018}
where the global step is Lagrangian and and the local step is
``almost''-Lagrangian. The essential steps described above are
summerized in Algorithm~\ref{algo:oneD}. Note that the steps to
reconstruct the multiscale basis are embarrassingly parallel and consist of small local problems.

\subsection{Basis Reconstruction in 2D} \label{s-2-5}

The basis reconstruction step in two dimensions is slightly
different. We separate the description of the procedure from the
reconstruction in one dimension to make the reader aware of
difficulties that arise when increasing the space dimension. Keeping
the difficulties in mind one can then easily generalize the procedure
even to 3D. We will comment on that again later. It is worth
mentioning that these difficulties do not need to arise in
non-conformal settings but since we agreed on a conformal method we
will be consistent here.

As the reader may have guessed to construct a $H^1$-conformal basis we
need to ensure that the reconstructed global basis is continuous
across coarse cell boundaries. This can be achieved by first looping
over all coarse edges of the traced back mesh and reconstructing the
solution at the previous time step $t^n$ with a basis representation
on each edge, i.e., we solve first an inverse problem of the
kind~(\ref{eq:2-4}). 
\begin{equation}
  \label{eq:2-100}
  \begin{split}
    \minimize_{\tilde\vphi_{\tilde \Gamma,i}\in C^0(\tilde \Gamma)} & 
    \quad  \norm{u^n|_{\tilde \Gamma} - \left(u_1^n \tilde \vphi_{\tilde \Gamma ,1} + u_2^n \tilde \vphi_{\tilde \Gamma ,2}\right)}{L^2(\tilde \Gamma)}^2 + \sum_i\alpha_i\R_i(\tilde\vphi_{\tilde \Gamma,i})\\
    \text{s.t.} & 
    \quad u_j^n = u^n(\ve{\tilde x}_{j}) \:, \quad \ve{\tilde x}_j \text{ is end point of (traced back) edge } \tilde \Gamma \\
    & \quad \vphi_{\tilde \Gamma ,i}(\ve{\tilde x}_{j}) = \delta_{ij} \:.\\
  \end{split}
\end{equation}
Note that the regularizer~(\ref{eq:2-5}) needs to
be replaced since the edge can be curved (it is unclear what
$\tilde\vphi^0_{K,i}$ in this case is). We use a harmonic prior for
the reconstructed basis 
\begin{equation}
  \label{eq:2-101}
   \R_i(\tilde\vphi_{\tilde K,i}) = \norm{\Delta_{g(\tilde\Gamma)}\tilde\vphi_{\tilde \Gamma,i}}{L^2(\tilde\Gamma)}^2
\end{equation}
with a low weight $\alpha_i$ in~(\ref{eq:2-4}). The operator
$\Delta_{g(\tilde \Gamma)}$ denotes the Laplace-Beltrami operator induced by the
standard Laplace operator with the trace topology of the respective
(traced back) edge $\tilde \Gamma$, i.e., $g(\tilde\Gamma)$ is the
metric tensor. We pass on providing details here.

The edge reconstructed basis then serves as boundary value for the
cell basis reconstruction. The optimization problem to solve on the
traced back cell $\tilde K$ then reads
\begin{equation}
  \label{eq:2-102}
  \begin{split}
    \minimize_{\tilde\vphi_{\tilde K,i}\in C^0(\tilde K)} & 
    \quad  \norm{u^n - \sum_{j=1}^3 u_j^n \tilde \vphi_{\tilde K ,j}}{L^2(\tilde K)}^2 + \sum_i\alpha_i\R_i(\tilde\vphi_{\tilde K,i})\\
    \text{s.t.} & 
    \quad u_j^n = u^n( \ve{\tilde x}_j ) \:, \quad \ve{\tilde x}_j \text{ is corner point of (traced back) cell } \tilde K \\
    & \quad \vphi_{\tilde K ,j}|_{\tilde \Gamma_l} = \vphi_{\tilde \Gamma ,j_l} \:.\\
  \end{split}
\end{equation}
However, the essential task is to ensure conformity of the global
basis by first reconstructing representations on all edges and then
inside the cells $\tilde K$.

In three dimensions it would be necessary to first reconstruct edge
then face and only then the interior representations. This sounds
potentially expensive but it is embarrassingly parallel since all
reconstructions are independent.

\subsection{Basis Propagation in 2D}\label{s-2-6}

Again, for didactical reasons we make the reader aware of differences
to the propagation of the basis in contrast to one spatial
dimension. As in Section~\ref{s-2-5} we need to preserve conformity of
the global basis. This can easily be done by fixing the boundary
values in the propagation step of the basis functions similar to the
propagation in 1D, see Section~\ref{s-2-4}.

However, this strategy can result in numerical errors from two
sources. Recall that the goal of our multiscale method is to represent
the local asymptotic structure of the solution correctly by putting
subgrid variations into the basis. This can be interpreted as adding a
non-polynomial corrector function in each coarse cell to a standard
FEM basis which changes the boundary conditions. These boundary
conditions are themselves subject to evolution and keeping them fixed
in time will not reflect any system intrinsic dynamics.

The first source of a numerical error is a resonance error.  This
error usually becomes dominant if the scale of oscillations in the
diffusion term is close to the scales resolved by the coarse
grid. This is well-documented in the literature on multiscale FEMs for
stationary elliptic problems that are usually not dominated by lower
order terms. Note, that in practical problems it is often unclear if a
scale separation exists or it can not be quantified which makes it
difficult to identify a resonance regime~\cite{Henning2013,HouWu1997}.

Another source for not capturing the asymptotic structure with fixed
boundary values even if there was a scale separation between coarse
mesh and diffusion oscillations (which can not be guaranteed) appears
in the conservative case~(\ref{eq:2-1a}). This is due to the reactive term 
\begin{equation}
  \label{eq:2-103}
  \left( \tilde\nabla \cdot\ve{\tilde c}_\delta\right) \vphi_{K,i}
\end{equation}
that appears also in~(\ref{eq:2-6b}). This term compensates
divergence, i.e., it triggers a reaction on the boundary wherever the
divergence of the background velocity does not vanish. Essentially
that means that this term is very local but can be quite strong, i.e.,
it is of the order of $\delta^{-1}$ if $\ve{c}$ is of order 1 but
oscillates on a scale of order $\delta$. This term must be taken into
account when evolving a reconstructed basis.

Fortunately, both errors can be dealt with by employing two
strategies. The first one is oversampling, i.e., the local
reconstruction and basis propagation on a coarse cell $K$ is performed on
a slightly larger domain than $K^{\mathrm{os}}\supset K$. After that
the basis is being truncated to the relevant domain $K$. Using this
strategy is generally possible but results in a non-conformal
technique and hence we will not use it
here~\cite{Henning2013}. Instead we will solve a reduced evolution
problem on the edges that predicts the boundary values and then evolve
the reconstructed interior basis for which we then know the boundary
values in time. The edge evolution strategy is in general only
necessary in the case of a missing scale separation for the diffusion
as explained above or in the conservative case. Note that both errors
do not appear in one dimension. 
We summarize this discussion below.

\vspace{0.25cm}

\paragraph{Strategy without Edge Evolution} In this case we keep the
boundary values of the basis that we reconstructed at the previous
time step fixed. Similarly to~(\ref{eq:2-6a}) this amounts to solving
the evolution problem
\begin{equation}
  \label{eq:2-104}
  \begin{split}
    \frac{\d}{\d t}\vphi_{K,i}
    & = \tilde\nabla\cdot\left( \ve{\tilde A}_\eps\tilde\nabla \vphi_{K,i} \right) \quad \text{in } \tilde K\times [ t^n,t^{n+1} ] \\
    \vphi_{K,i}(\cdot,t)|_{\tilde \Gamma_l} 
    & = \tilde\vphi_{\tilde \Gamma_l,i} \:, \quad t\in [ t^n,t^{n+1} ] \\
    \vphi_{K,i}(\ve{\tilde x},t^n) 
    & = \tilde\vphi_{\tilde K,i}(\ve{\tilde x})
  \end{split}
\end{equation}
for the $i$-th basis function. 

\vspace{0.25cm}

\removelatexerror
\begin{algorithm}[t!]
  \setstretch{1}
  \RestyleAlgo{ruled}
  \SetAlgoLined
  \LinesNumbered
  \SetKwInOut{Input}{input}
  \SetKwInOut{Output}{output}
  \SetKwFor{ForP}{for}{do in parallel}{end}
  
  \Input{Problem parameters for~(\ref{eq:2-1a}) or~(\ref{eq:2-1b}),
    $H$, $h$, $\delta t$}
  
  \Output{Weights $\ve u^H(t_j)$ of multiscale solution and set of
    multiscale basis functions $\vphi_i^{H,\mathrm{ms}}$}
  
  \Begin{      
    Initialize a coarse mesh $\T_H$\;
    
    On each cell $K\in\T_H$ initialize a fine mesh $\T_h^K$\;

    \tcc{We need to reconstruct a basis at time $t^0$. Note that no
      trace back and no propagation are needed.}  
    
    \ForP{$K\in\T_H$}{
      
      Reconstruct the optimal basis representation of
      $\left. u_0(x)\right|_K$ according to~(\ref{eq:2-4})\;

    }

    \tcc{Now the time stepping starts. We compute the solution at $t^{n+1}$.}
    \For{$n=0$ \KwTo $n\leq N_{\mathrm{steps}}$}{
      
      \ForP{$K\in\T_H$}{ 
        
        Trace back each node in $K$ one time step from $t^{n+1}$ to
        $t^n$ according to~(\ref{eq:2-2})\;
        
        Reconstruct the boundary condition of optimal basis 
        representation of $\left. u^n(x)\right|_{\tilde K}$ according
        to~(\ref{eq:2-100}), see Section~\ref{s-2-5}\;

        Reconstruct the optimal basis representation of
        $\left. u^n(x)\right|_{\tilde K}$ according
        to~(\ref{eq:2-102}), see Section~\ref{s-2-5}\;

        \uIf{Problem is not in conservation form}{
          Propagate the optimal basis forward onto $K$ according
          to~(\ref{eq:2-104})\;
        }
        \Else{
          Propagate the boundary conditions of the optimal basis 
          forward onto $K$ according to~(\ref{eq:2-105})\; 

          Propagate the optimal basis forward onto $K$ according
          to~(\ref{eq:2-106})\; }
        }        

      Assemble the global (coarse) system matrices for~(\ref{eq:2-13}) \;

      Make a global backward Euler time step using~(\ref{eq:2-13})\;

    }

    Postprocess the solution\;

    Return\;

  }
  
  \caption{Sketch of modified reconstruction algorithm for a 2D
    problem. Implicit version using backward Euler.}
  \label{algo:twoD}
\end{algorithm}

\paragraph{Strategy with Edge Evolution} This method requires a
prediction of the boundary values. The prediction is done by solving a
reduced evolution problem
\begin{equation}
  \label{eq:2-105}
  \begin{split}
    \frac{\d}{\d t}\vphi_{\Gamma,i} + \left( \tilde\nabla \cdot\ve{\tilde c}_\delta^{\mathrm{red}}\right) \vphi_{\Gamma,i}
    & = \tilde\nabla\cdot\left( \ve{\tilde A}_\eps^{\mathrm{red}}\tilde\nabla \vphi_{\Gamma,i} \right) \quad \text{on } \tilde \Gamma\times [ t^n,t^{n+1} ] \\
    \vphi_{K,i}(x_j,t)
    & = \delta_{ij} \:, \quad t\in [ t^n,t^{n+1} ] \:,\:\ve{\tilde x}_j \text{ is end point of (traced back) edge } \tilde \Gamma \\
    \vphi_{\Gamma,i}(\ve{\tilde \Gamma},t^n) 
    & = \tilde\vphi_{\tilde \Gamma,i}(\ve{\tilde x})
  \end{split}
\end{equation}
on each edge $\tilde \Gamma$ of $\tilde K$. The reduced quantities
$\ve{\tilde c}_\delta^{\mathrm{red}}$ and
$\ve{\tilde A}_\eps^{\mathrm{red}}$ only contain the tangential parts
of their counterparts $\ve{\tilde c}_\delta$ and $\ve{\tilde A}_\eps$.
Once this evolution problem is solved for each edge $\tilde \Gamma$
the boundary values for the full evolution problem on $\tilde K$ are
known and one can solve
\begin{equation}
  \label{eq:2-106}
  \begin{split}
    \frac{\d}{\d t}\vphi_{K,i} + \left( \tilde\nabla \cdot\ve{\tilde c}_\delta\right) \vphi_{\Gamma,i}
    & = \tilde\nabla\cdot\left( \ve{\tilde A}_\eps\tilde\nabla \vphi_{K,i} \right) \quad \text{in } \tilde K\times [ t^n,t^{n+1} ] \\
    \vphi_{K,i}(\cdot,t)|_{\tilde \Gamma_l} 
    & = \tilde\vphi_{\tilde \Gamma_l,i}(\cdot,t) \:, \quad t\in [ t^n,t^{n+1} ] \\
    \vphi_{K,i}(\ve{\tilde x},t^n) 
    & = \tilde\vphi_{\tilde K,i}(\ve{\tilde x})
  \end{split}
\end{equation}
to get the $i$-th basis.

\begin{wrapfigure}{l}{0.4\textwidth}
  \vspace{-0pt}
  \begin{center}
    \includegraphics[width=0.39\textwidth]{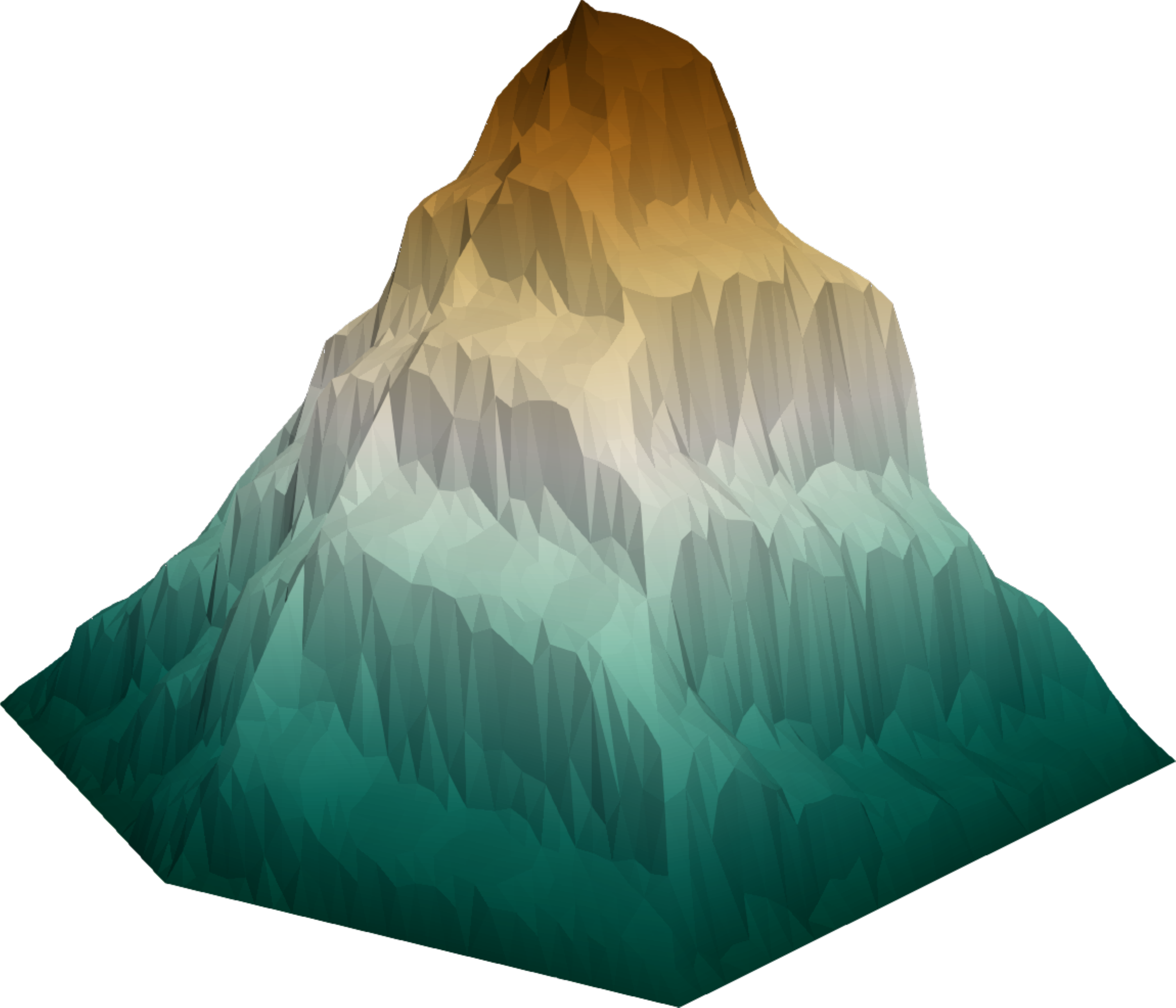}
  \end{center}
  \vspace{0pt}
  \caption{Illustration of a reconstructed multiscale basis in 2D.}
  \vspace{-0pt}
  \label{fig-2d_basis}
\end{wrapfigure} In both cases note again that the evolution
equations~(\ref{eq:2-104}) and~(\ref{eq:2-106}) are solved on
$\tilde K$. The end state $\vphi_{K,i}(\ve{\tilde x},t^{n+1})$ on
$\tilde K$ can again be mapped onto the Eulerian element $K\in\T_H$ to
obtain the desired basis function
$\vphi_{K,i}(\ve{x},t^{n+1})\sim \vphi_{K,i}^{n+1}(\ve{x})$ at the
next time step.

The described modifications for 2D problems are summarized in
Algorithm~\ref{algo:twoD}.
\section{Numerical Examples in 1D}\label{s-3}

We will show several 1D examples in a non-conservative and
conservative setting according to~(\ref{eq:2-1a}) and~(\ref{eq:2-1b}),
respectively. For all 1D tests we use a Gaussian
\begin{equation}
  \label{eq:2-14}
  u_0(x) = \frac{1}{\sigma\sqrt{2\pi}}\exp{-\frac{(x-\mu)^2}{2\sigma^2}} \:.
\end{equation}
with variance $\sigma=0.1$ centered in the middle of the domain
$\tor^1$, i.e., $\mu=0.5$. The end time is set to $T=1$ with a time
step $\delta t = 1/300$. We show our semi-Lagrangian multiscale
reconstruction method (SLMsR) with a coarse resolution $H=10^{-1}$ in
comparison to a standard FEM with the same resolution and high order
quadrature. As a reference method we choose a high-resolution standard
FEM with $h_{\mathrm{ref}}=10^{-3}$. For the multiscale method we
choose a fine mesh $\T_h^K$ with $h=10^{-2}$ in each coarse cell
$K\in\T_H$.

\paragraph{Test 1} We start with four tests showing the capability of
the SLMsR to capture subgrid variations correctly. Note that the
coarse standard FEM has as many cells the the SLMsR has coarse cells
and that it does not capture subgrid variations in the following
tests. The resolution for the reference solution resolves all subgrid
variations but the reader should keep in mind that practical
applications do not allow the application of high-resolution
methods. The coefficients
\begin{equation}
  \label{eq:2-15}
  \begin{split}
    c_\delta(x,t) & = \frac{1}{4} + \frac{1}{2}\cos(10\pi x) + \frac{1}{4}\cos(74\pi x) + \frac{3}{20}\cos(196\pi x) \\
    A_\eps(x,t) & = 10^{-3} + 9\cdot 10^{-4}\cos(10\pi t) \cos(86\pi x) \\
  \end{split}
\end{equation}
and 
\begin{equation}
  \label{eq:2-16}
    \begin{split}
      c_\delta(x,t) & = \frac{1}{2}\cos(2\pi t) + \frac{1}{4}\cos(6\pi t)\cos(8\pi x) + \frac{1}{8}\cos(4\pi t)\cos(62\pi x) + \frac{1}{8}\cos(150\pi x) \\
      A_\eps(x,t) & = 10^{-3} + 9\cdot 10^{-4}\cos(10\pi t) \cos(86\pi x) \\
  \end{split}
\end{equation}
are chosen for the non-conservative equation~(\ref{eq:2-1a}) and the
coefficients
\begin{equation}
  \label{eq:2-17}
    \begin{split}
      c_\delta(x,t) & = \frac{1}{2} + \frac{1}{8}\cos(8\pi x) + \frac{1}{8}\cos(62\pi x) + \frac{1}{8}\cos(150\pi x) \\
      A_\eps(x,t) & = 10^{-2} + 9\cdot 10^{-3}\cos(10\pi t) \cos(86\pi x) \\
  \end{split}
\end{equation}
and
\begin{equation}
  \label{eq:2-18}
    \begin{split}
      c_\delta(x,t) & = \frac{3}{4} + \frac{1}{2}\cos(8\pi x) + \frac{1}{4}\cos(62\pi x) + \frac{1}{10}\cos(150\pi x) \\
      A_\eps(x,t) & = 10^{-2} + 9\cdot 10^{-3}\cos(10\pi t) \cos(86\pi x) \\
  \end{split}
\end{equation}
\clearpage
\begin{figure}[t!]
  \vspace{0pt}
  \begin{center}
    \includegraphics[width=1.0\textwidth]{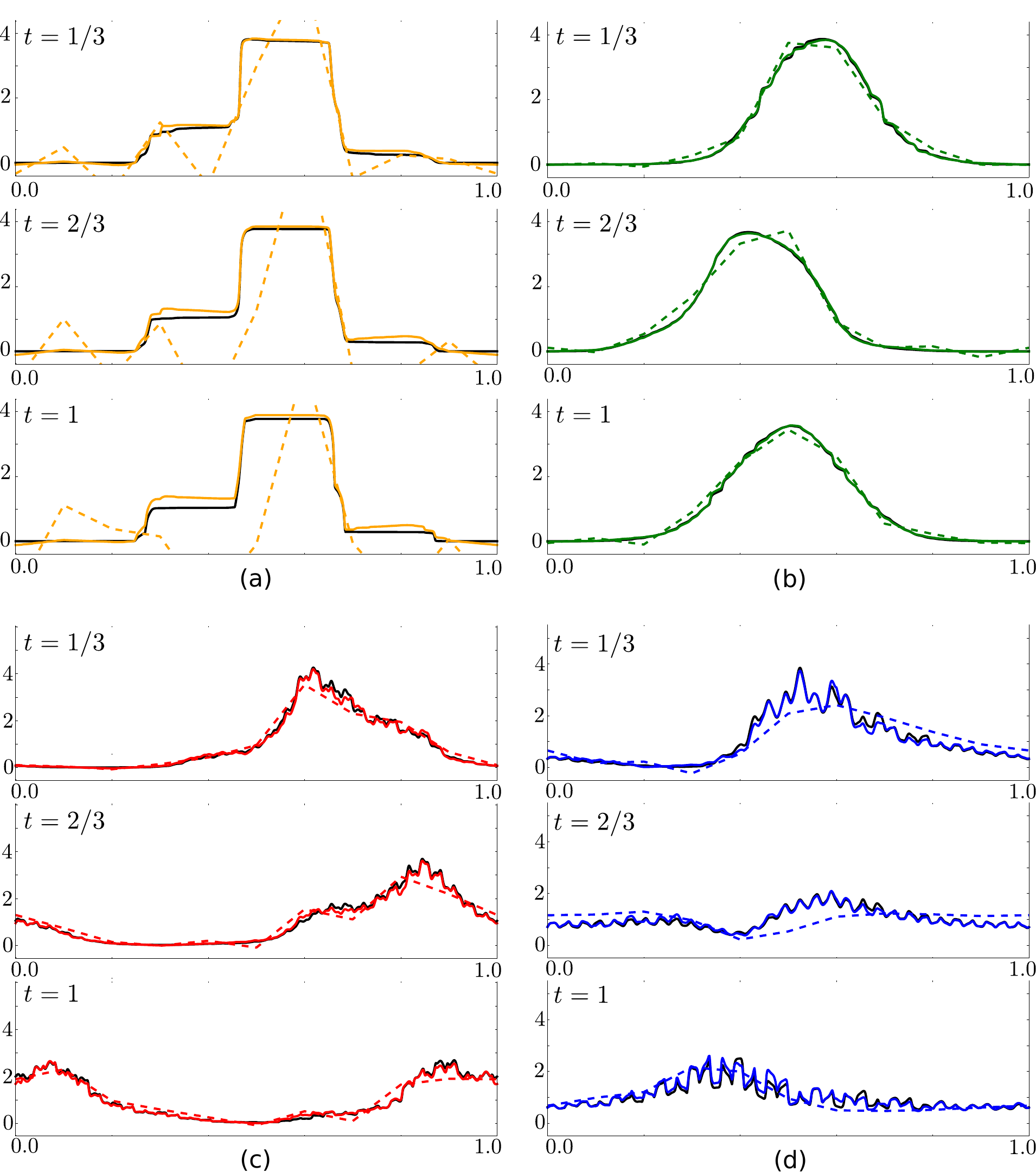}
  \end{center}
  \vspace{0pt}
  \caption{Snapshots of the solution at $t=1/3$, $t=2/3$ and
    $T=1$. The colored dashed lines show the solution of the standard
    FEM, the colored line shows the SLMsR. The reference solution is
    shown in black. \textbf{(a)} Non-conservative
    equation~(\ref{eq:2-1a})
    Coefficients~(\ref{eq:2-15}). \textbf{(b)} Non-conservative
    equation~(\ref{eq:2-1a})
    Coefficients~(\ref{eq:2-16}). \textbf{(c)} Conservative
    equation~(\ref{eq:2-1b})
    Coefficients~(\ref{eq:2-17}). \textbf{(d)} Conservative
    equation~(\ref{eq:2-1b}) Coefficients~(\ref{eq:2-18}).}
  \vspace{0pt}
  \label{fig-1d_examples_series}
\end{figure}
\clearpage

\noindent conservative equation~(\ref{eq:2-1b}). The latter one is
numerically more difficult when it comes to capturing fine-scale
variations, i.e., if $c_\delta(x)\sim f(x/\delta)$ then
$\frac{\d}{\d x}c_\delta(x)\sim\delta^{-1}f'(x/\delta)$. So if $f$ and
$f'$ are of the same order like in one term of a Fourier expansion of
$c_\delta$ then one can expect very steep slopes in the solution. The
results of the tests are shown in Figure~\ref{fig-1d_examples_series}
and the corresponding errors in
Figure~\ref{fig-1d_examples_series_error}.

\paragraph{Test 2a} It is an important question how the SLMsR behaves
in different regimes of data. For the non-conservative case we show
two tests.

The first test demonstrates how the SLMsR behaves when all
coefficients of the equation are resolved by the SLMsR, i.e.,
$Hh\leq\eps,\delta$ but not by the low resolution FEM that has the
same resolution as the SLMsR's coarse resolution, i.e.,
$H\gg\eps,\delta$. For the coefficients we chose
\begin{equation}
  \label{eq:2-19}
    \begin{split}
      c_\delta(x,t) & = \frac{1}{2} + \frac{1}{4}\cos(8\pi x) + \frac{1}{8}\cos(196\pi x) + \frac{1}{16}\cos(210\pi x) \\
      A_\eps(x,t) & = 10^{-3} + 9\cdot 10^{-4}\cos(174\pi x) \:. \\
  \end{split}
\end{equation}
The different spatial resolutions were taken as
$H=\frac{1}{8},\frac{1}{32},\frac{1}{128},\frac{1}{512}$ while the
number of fine cells in each coarse cell was fixed to $n_f=64$. The
reference solution was computed with $h_{\mathrm{ref}}=2^{-11}$ and
the time step for all methods was taken as $\delta t=10^{-2}$. 

\begin{figure}[t!]
  \vspace{0pt}
  \begin{center}
    \includegraphics[width=1.0\textwidth]{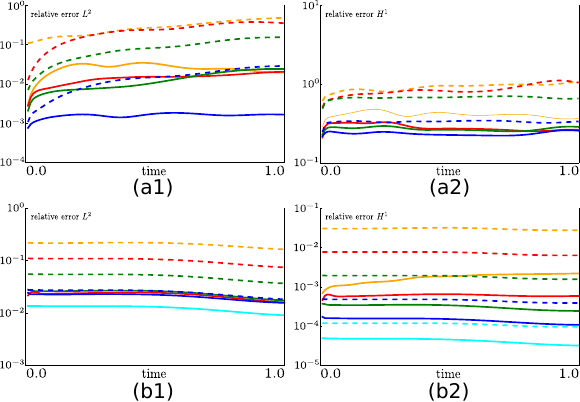}
  \end{center}
  \vspace{0pt}
  \caption{Relative errors over time for the unresolved regime of test
    problem~(\ref{eq:2-19}) in $L^2(\tor^1)$ are shown in
    \textbf{(a1)} and errors in $H^1(\tor^1)$ are shown in
    \textbf{(a2)}. The erros of the numerically resolved regime of
    problem~(\ref{eq:2-20}) in $L^2(\tor^1)$ are shown in
    \textbf{(b1)} and errors in $H^1(\tor^1)$ are shown in
    \textbf{(b2)}. The respective reference solutions were computed
    with $h_{\mathrm{ref}}=2^{-11}$ and for all experiments we used
    $\delta t=10^{-2}$. The dashed lines show the error of the
    standard FEM, the full line shows the error of the SLMsR. The
    coarse resolutions were chosen as $H=1/8$ (yellow), $H=1/32$
    (red), $H=1/128$ (green), $H=1/256$ (blue), and $H=1/512$ (cyan,
    only second row). All errors are shown in a logarithmic scale.}
  \vspace{0pt}
  \label{fig-1d_convergence_test}
\end{figure}

Error plots of the results are shown in the first row of
Figure~\ref{fig-1d_convergence_test}. The plots indeed show that the
SLMsR captures the reference solution much more accurately than the
standard FEM in both $L^2$ and $H^1$ even for small $H$. Only when
decreasing $H$ to a resolution that resolves all coefficients the FEM
is able to represent the solution correctly and starts converging. The
reader should keep in mind that in real world applications the
standard FEM will be too expensive to compute while the SLMsR allows
for massive parallelization. This is the regime that is of practical
interest. Increasing the fine resolution in each coarse cell on the
other hand does not improve the the accuracy of the SLMsR. It
increases the size of the subgrid problems for the basis functions
though.

\paragraph{Test 2b} The second test is to show that the SLMsR
essentially starts behaving like a standard FEM if $H\ll \eps,\delta$.
To demonstrate that we took low frequency coefficients
\begin{equation}
  \label{eq:2-20}
    \begin{split}
      c_\delta(x,t) & = \frac{1}{2} + \frac{1}{4}\cos(6\pi x) + \frac{1}{8}\cos(10\pi x) + \frac{1}{16}\cos(14\pi x) \\
      A_\eps(x,t) & = 10^{-3} + 9\cdot 10^{-4}\cos(16\pi x) \:. \\
  \end{split}
\end{equation}
and refined with the sequence
$H=\frac{1}{16},\frac{1}{32},\frac{1}{64},\frac{1}{128}$ while we we
fixed the number of fine cells in each coarse cell to $n_f=32$. Note
that we start with a coarse resolution regime that almost resolves the
data. The error plots for the SLMsR and the FEM indeed indicate that
as $H$ decreases the SLMsR does not dramatically increase its accuracy
while the FEM with linear basis functions increases its accuracy
according to the standard estimates. This can be observed in the
second row of Figure~\ref{fig-1d_convergence_test}. The relative
errors at $t=1/2$ and $t=1$ for both test problems~(\ref{eq:2-19})
and~(\ref{eq:2-20}) are summarized in
Tables~\ref{tabl-1d_convergence_unresolved}
and~\ref{tabl-1d_convergence_resolved} in the appendix together with
the estimated order of convergence (EOC) in space. We recall that the
EOC can be computed as follows: Let $e_H:\real_+^n\rightarrow\real_+$
be an (error) function that maps a high dimensional vector to a
positive real for each $H>0$. Then the EOC w.r.t. $H$ can then be
computed as
\begin{equation}
  \label{eq:2-21}
  \mathrm{EOC} = \frac{\log(e_{H/q}) - \log(e_H)}{\log(q)} \:.
\end{equation}

\clearpage
\begin{figure}[t!]
  \vspace{0pt}
  \begin{center}
    \includegraphics[width=1.0\textwidth]{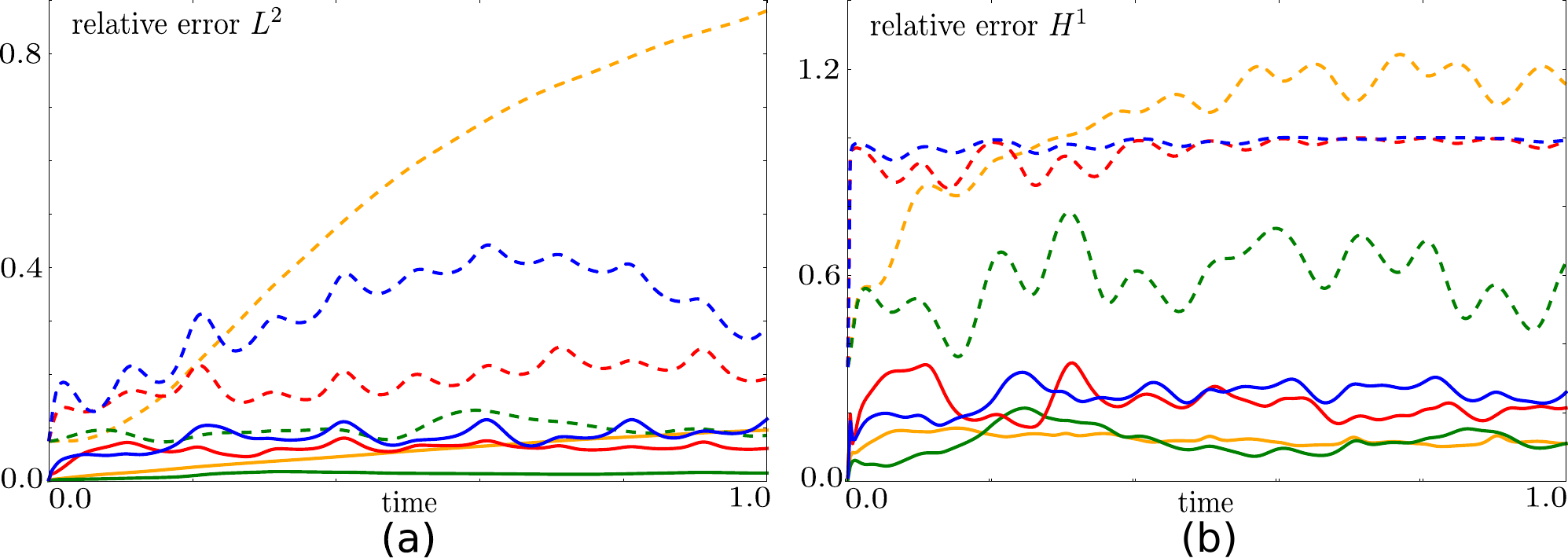}
  \end{center}
  \vspace{0pt}
  \caption{Relative errors over time in \textbf{(a)} $L^2(\tor^1)$ and
    \textbf{(b)} $H^1(\tor^1)$ to the reference solution of the tests shown in
    Figure~\ref{fig-1d_examples_series}. The dashed lines show the
    error of the standard FEM, the full line shows the error of the
    SLMsR. Color codes are the same as in
    Figure~\ref{fig-1d_examples_series}.}
  \vspace{0pt}
  \label{fig-1d_examples_series_error}
\end{figure}
\clearpage

\noindent Similar results in terms of accuracy can be obtained in case
of a conservative equation. We pass on showing this here.

\begin{figure}[t!]
  \vspace{0pt}
  \begin{center}
    \includegraphics[width=1.0\textwidth]{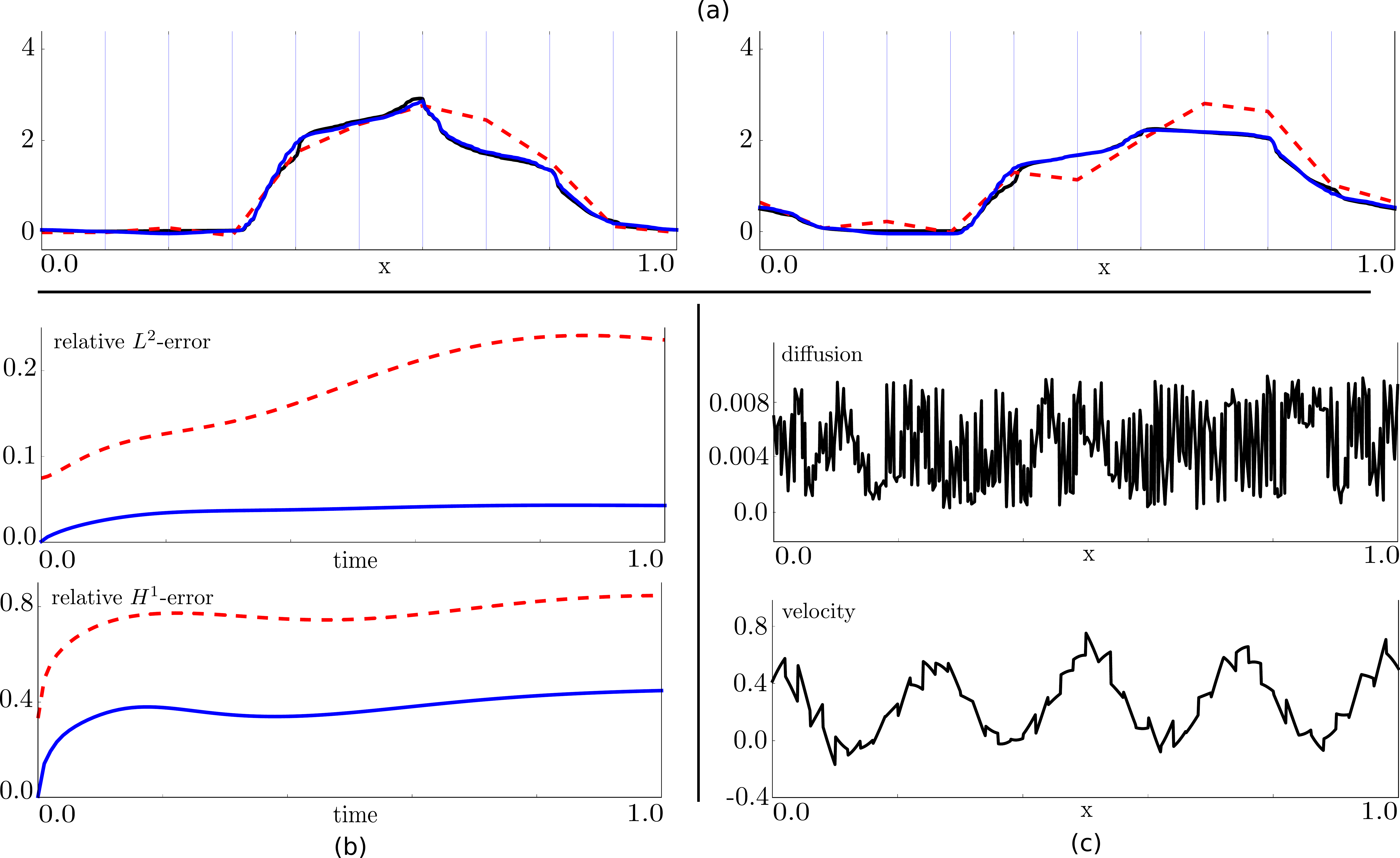}
  \end{center}
  \vspace{0pt}
  \caption{Comparison of SLMsR and standard FEM for randomly generated
    (but fixed) coefficient functions. \textbf{(a)} Snapshots of the
    solution at $t=1/2$ and $t=1$. Solid black lines show the reference
    solution, dashed red lines show the standard solution and solid blue
     lines show the SLMsR. \textbf{(b)} Relative error plots for
    $H^1$ and $L^2$. \textbf{(c)} Coefficient functions: the diffusion
    coefficient (upper plot) was generated from a uniform distribution and
    scale to have minumum $10^{-5}$ and maximum $10^{-2}$. The velocity
    coeffcient (lower plot) we chose to be a smooth function disturbed by
    Gaussian noise with mean zero and variance $0.1$.}
  \vspace{0pt}
  \label{fig-1d_random}
\end{figure}

\paragraph{Test 3} This test shows an example where both diffusion
and background velocity are generated randomly. We intend to show an
example of the SLMsR behaves when data is involved that does not
exhibit a clear scale separation. For this we initially generate fixed
mesh based functions with random nodal coefficients. In each mesh cell
the functions are interpolated linearly. Note that this is not to
simulate a sampled stochastic process. We simply intend not to create
any scale or symmetry bias when constructing coefficient
functions. The results look appealing and show a clear advantage of
the SLMsR, see Figure~\ref{fig-1d_random}.
\section{Numerical Examples in 2D}\label{s-4}

This section is to experimentally demonstrate that the SLMsR can
handle conservative and non-conservative advection-diffusion equations
with dominant advection term in higher dimensions. All tests are being
carried out on the torus $\tor^2$ (periodic unit square) in the time
interval $t\in [0,1]$ and use the same initial condition. As initial
value we chose a normalized super-position of two non-isotropic
Gaussians
\begin{equation}
  \label{eq:4-1}
  u_0(\ve x) = \frac{1}{2\sqrt{(2\pi)^2\det (\ve M)}} 
  \sum_{i=1}^2
  \exp\left\lbrace 
    -\frac{1}{2}(\ve x-\boldsymbol \mu_i)^T\ve M^{-1}(\ve x-\boldsymbol \mu_i) 
  \right\rbrace 
\end{equation}
where
\begin{equation}
  \label{eq:4-2}
  \ve M = 
  \begin{bmatrix}
    \frac{3}{100} & 0 \\
    0 & \frac{3}{100} \\
  \end{bmatrix}
  \quad \text{and} \quad
  \boldsymbol\mu_i = 
  \begin{bmatrix}
    \frac{i}{3} \:, & \frac{1}{2}
  \end{bmatrix}^T \:.
\end{equation}
All tests of the SLMsR are are perfomed on a coarse unstructured
uniform triangular Delaunay mesh with $n_c=62$ coarse cells, i.e., for
our triangulation $H\sim 0.3$ (mean diameter of circumcircle of a
cell). We compare the SLMsR to a standard low resolution FEM with the
same resolution and to a standard high resolution FEM with
approximately $n_f=63K$ cells. To get a fine mesh on each coarse cell
of the SLMsR we created a triangulation such that the sum of all fine
cells over all coarse cells is \begin{wrapfigure}[19]{r}{0.42\textwidth}
  \vspace{0pt}
  \begin{center}
    \includegraphics[width=0.42\textwidth]{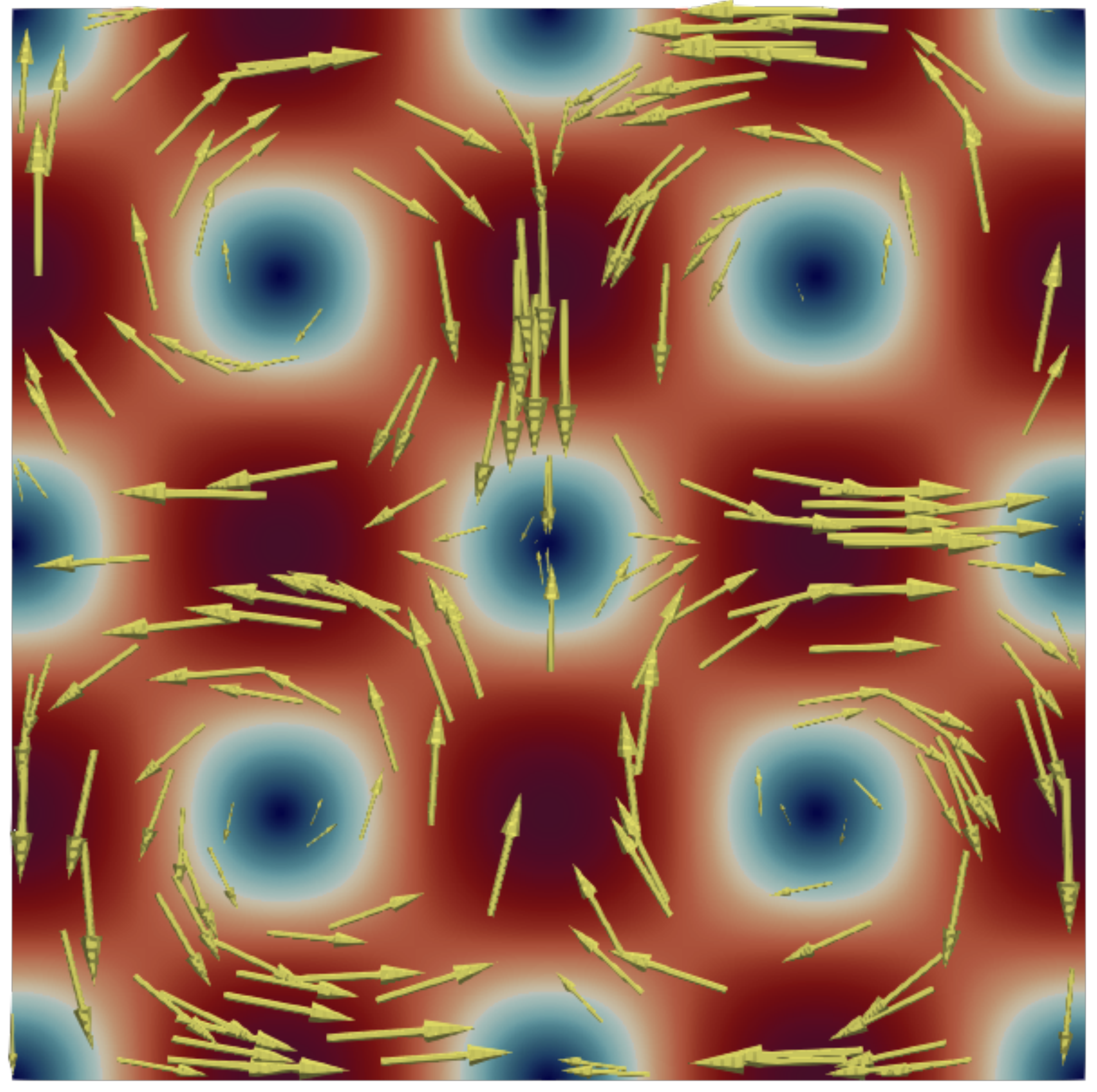}
  \end{center}
  \vspace{0pt}
  \caption{Background velocity for \emph{Test 1} and \emph{Test
      3}. Four vortices moving through the domain from left to right
    and come back to their starting points at $T=1$.}
  \vspace{0pt}
  \label{fig-2d_GaussSolenoidalVel}
\end{wrapfigure}
approximately $n_f$ to get a fair comparison of the SLMsR to the low
resolution standard FEM with respect to the reference solution that
resolves all coefficients involved.

\paragraph{Test 1} Here we test our multiscale reconstruction method
with a solenoidal field $\ve c_\delta$ described by the stream
function
\begin{equation}
  \label{eq:4-3}
  \psi(\ve x, t) = \sin(2\pi(x_1-t))\sin(2\pi x_2)
\end{equation}
so that
\begin{multline}
  \label{eq:4-4}
  \ve c_{\delta}(\ve x, t) = \nabla^\perp\psi \\ = 
  2\pi
  \begin{bmatrix}
    \sin(2\pi(x_1-t))\cos(2\pi x_2) \\ -\cos(2\pi(x_1-t))\sin(2\pi x_2) \\
  \end{bmatrix} \:.
\end{multline}
This background velocity desribes four vortices moving in time through
the (periodic) domain from left to right and get back to their
starting point at $T=1$. Note that this velocity field involves scales
that are resolved by the coarse mesh as well as scales that are not
resolved, see Figure~\ref{fig-2d_GaussSolenoidalVel}. Note that since
$\DIV \ve c_\delta=0$ equation~(\ref{eq:2-1a}) and~(\ref{eq:2-1b}) are
(analytically) identical and hence we only solve~(\ref{eq:2-1a}).

\begin{figure}[b!]
  \vspace{-0pt}
  \begin{center}
    \includegraphics[width=1.0\textwidth]{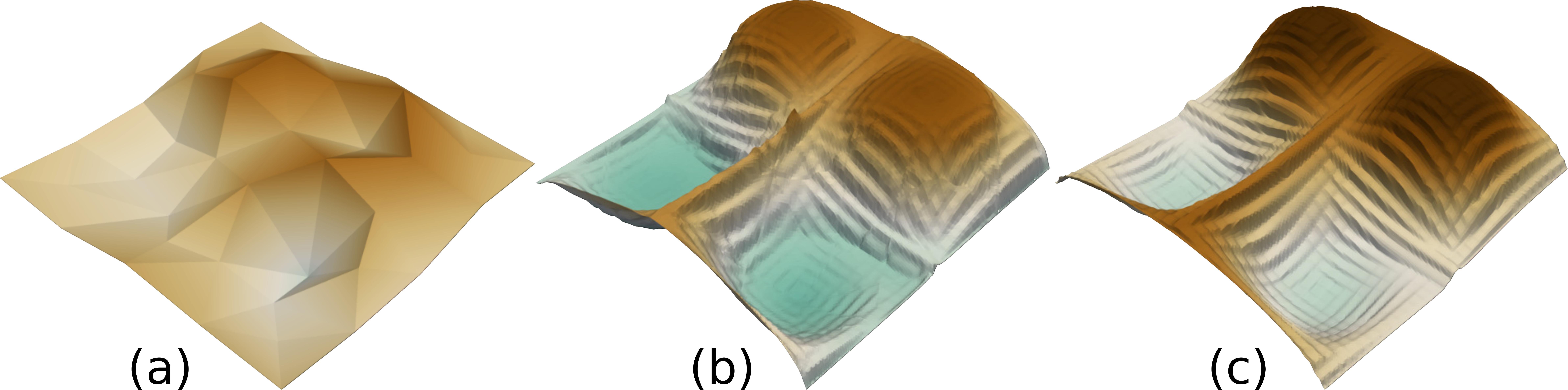}
  \end{center}
  \vspace{0pt}
  \caption{Snapshots of the solution for \emph{Test 1} at time $T=1$
    for \textbf{(a)} the low resolution standard FEM, \textbf{(b)} the
    SLMsR and \textbf{(c)} the reference solution.}
  \vspace{0pt}
  \label{fig-2d_GaussSolenoidal}
\end{figure}

\begin{figure}[t!]
  \vspace{-0pt}
  \begin{center}
    \includegraphics[width=1.0\textwidth]{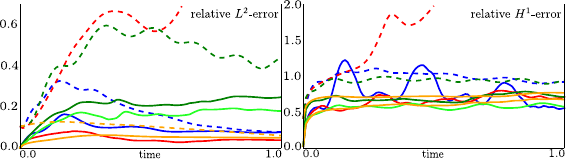}
  \end{center}
  \vspace{0pt}
  \caption{Relative errors in $L^2(\tor)$ and $H^1(\tor)$ over time
    for all test cases of Section~\ref{s-4}. The SLMsR (solid lines)
    in general shows lower errors than the low resolution standard FEM
    (dashed lines). The color codes are as follows: \emph{Test 1} is
    shown in blue, \emph{Test 2} is shown in red, \emph{Test 3} is
    shown in green where the dark green solid line shows the error
    without edge evolution and the light green solid line shows the
    error with edge evolution according
    to~(\ref{eq:2-105}). \emph{Test 4} is shown in yellow.}
  \vspace{0pt}
  \label{fig-2d_error}
\end{figure}

The diffusion tensor is chosen to be
\begin{equation}
  \label{eq:4-5}
  \ve A_\eps(\ve x, t) = \frac{1}{100}\id_2 - 0.9999
  \begin{bmatrix}
    \sin(60\pi x_1) & 0 \\
    0 & \sin(60\pi x_2) \\
  \end{bmatrix} \:.
\end{equation}
Note that in this case advection dominance is a local property and
P\'{e}clet numbers are ranging from $\peclet = 0$ to
$\peclet\sim 6\cdot 10^6$. Snapshots of the solutions are shown in
Figure~\ref{fig-2d_GaussSolenoidal}. \begin{wrapfigure}[26]{l}{0.39\textwidth}
  \vspace{0pt}
  \begin{center}
    \includegraphics[width=0.39\textwidth]{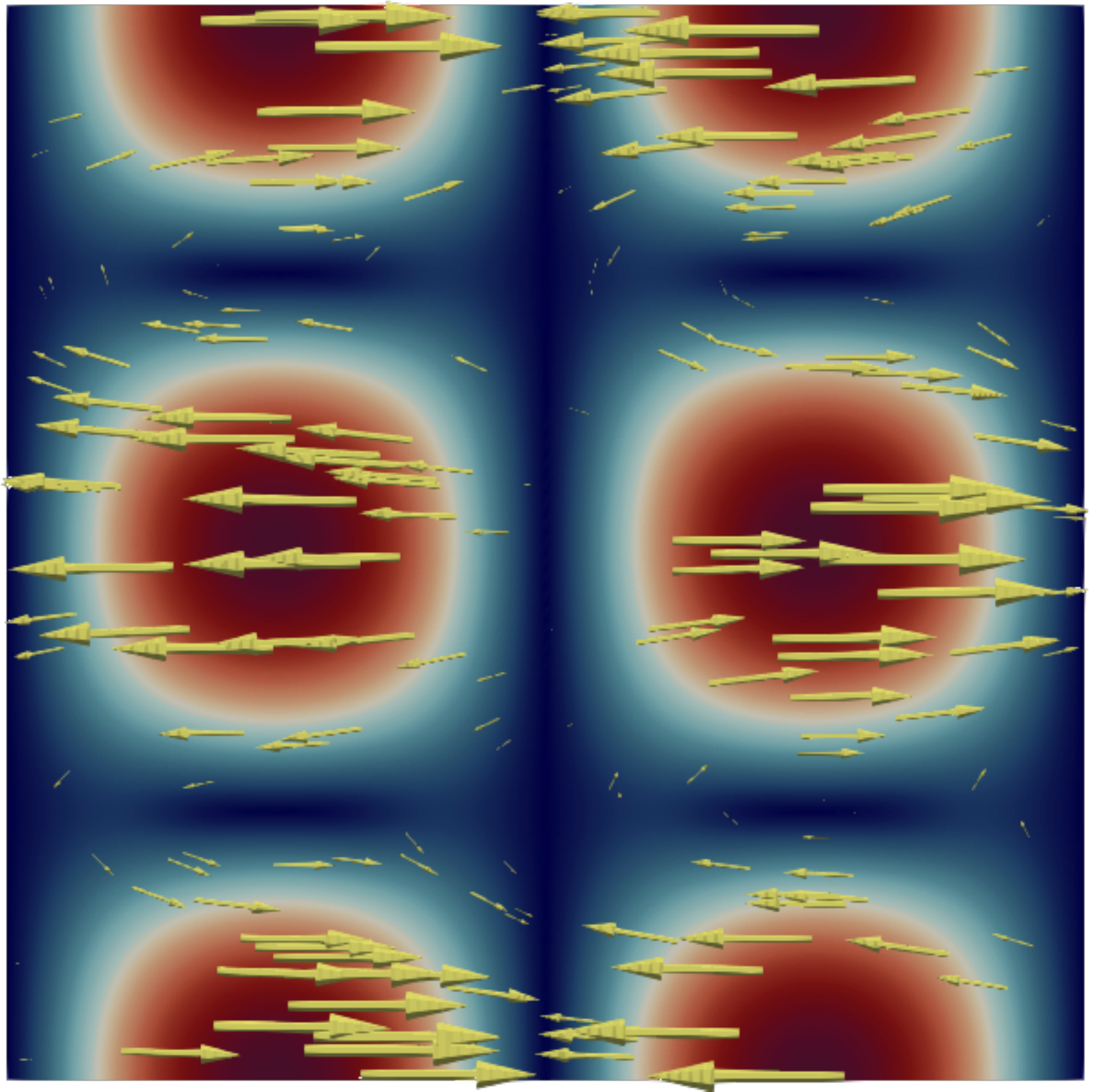}
    \includegraphics[width=0.39\textwidth]{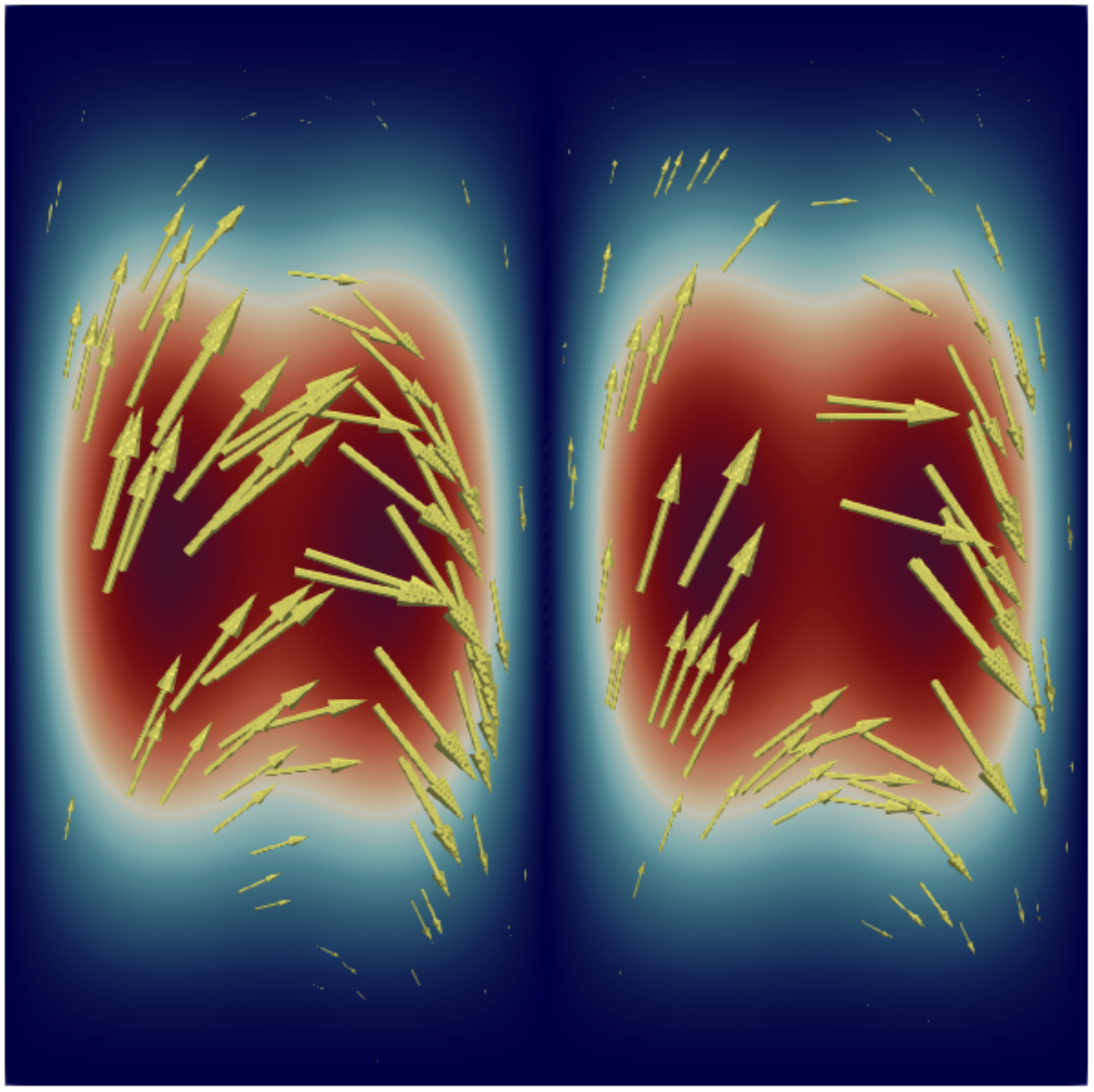}
  \end{center}
  \vspace{0pt}
  \caption{Background velocities for \emph{Test 2a} and \emph{Test 2b}
    given by~(\ref{eq:4-6}) and~(\ref{eq:4-7}), respectively. Both
    fields are divergent.}
  \vspace{0pt}
  \label{fig-2d_GaussDivVel}
\end{wrapfigure} It can
be observed that the low resolution FEM does not capture the effective
solution well since it diffuses too strongly while the SLMsR
reasonably caputeres the effective behavior of the solution and even
the fine scale structure. This can be seen also in the error plots,
Figure~\ref{fig-2d_error}.

\begin{figure}[t!]
  \vspace{-0pt}
  \begin{center}
    \includegraphics[width=1.0\textwidth]{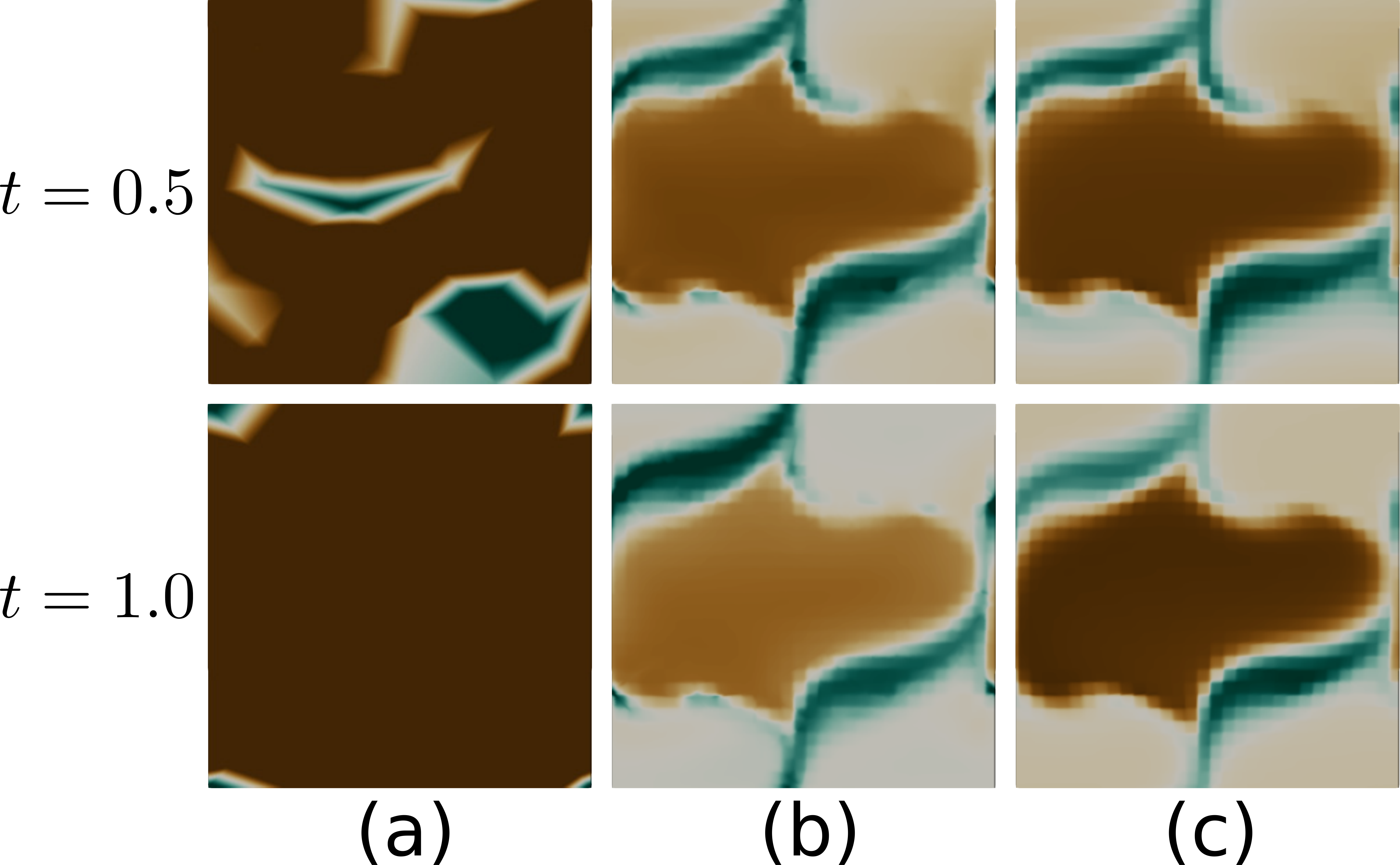}
  \end{center}
  \vspace{0pt}
  \caption{Comparison of \textbf{(a)} low resolution FEM and
    \textbf{(b)} SLMsR to \textbf{(c)} the reference solution of
    \emph{Test 2a} at two time instances.}
  \vspace{0pt}
  \label{fig-2d_GaussDiv}
\end{figure}

\paragraph{Test 2a} For this test we use two divergent background
velocities and hence we split it into two parts to distiguish between
the diffusive transport problem~(\ref{eq:2-1a}) and the conservation
law~(\ref{eq:2-1b}). We start with showing how the SLMsR behaves on
equation~(\ref{eq:2-1a}). The velocity field is a modification
of~(\ref{eq:4-4}) and is given by \vspace{0.1cm}
\begin{multline}
  \label{eq:4-6}
  \ve c_\delta(\ve x, t) = 
  \begin{bmatrix}
    \cos(2\pi t) & \sin(2\pi t) \\
    -\sin(2\pi t) & \cos(2\pi t) \\
  \end{bmatrix}
  \cdot
  \\
  \begin{bmatrix}
    2\pi\sin(2\pi(x_1-t))\cos(2\pi x_2) \\ -\cos(2\pi(x_1-t))\sin(2\pi x_2)\sin(2\pi x_1) \\
  \end{bmatrix} \:.
\end{multline}
while the diffusion tensor is also given by~(\ref{eq:4-5}). Here the
standard solution does not converge to any reasonable approximation of
the effective solution and no valuable understanding of the dynamics
can be drawn from it. The SLMsR shows a suprisingly good
approximation of the reference solution, see
Figure~\ref{fig-2d_GaussDiv}.

\paragraph{Test 2b} Here we solve equation~(\ref{eq:2-1b}), i.e., a
conservation problem. The divergent velocity field for this test
describes regions of fast and slow flow with two separatrices across
which there is no flow moving in time from left to right once through
the domain during the time interval. The field is given by
\begin{equation}
  \label{eq:4-7}
  \ve c_\delta(\ve x, t) = \frac{2\pi}{5}
  \begin{bmatrix}
    \sin(2\pi(x_1-t))^2 \cos(\pi(x_2-0.5))^2 \\ 2\sin(2\pi(x_1-t)) \cos(2\pi (x_1-t)) \cos(\pi(x_2-0.5))^2 \:.
  \end{bmatrix}
\end{equation}
Both velocity fields for \emph{Test 2a} and \emph{Test 2b} are shown
in Figure~\ref{fig-2d_GaussDivVel}. The diffusion tensor is again
given by~(\ref{eq:4-5}). The difference to the previous test is that
we have quite a large additional reaction term to handle. This will be
taken care of in the basis representation as described in
Sections~\ref{s-2-5} and~\ref{s-2-6} as well as in the gloabl weak
form, see equation~(\ref{eq:2-10b}).

\begin{figure}[t!]
  \vspace{-0pt}
  \begin{center}
    \includegraphics[width=1.0\textwidth]{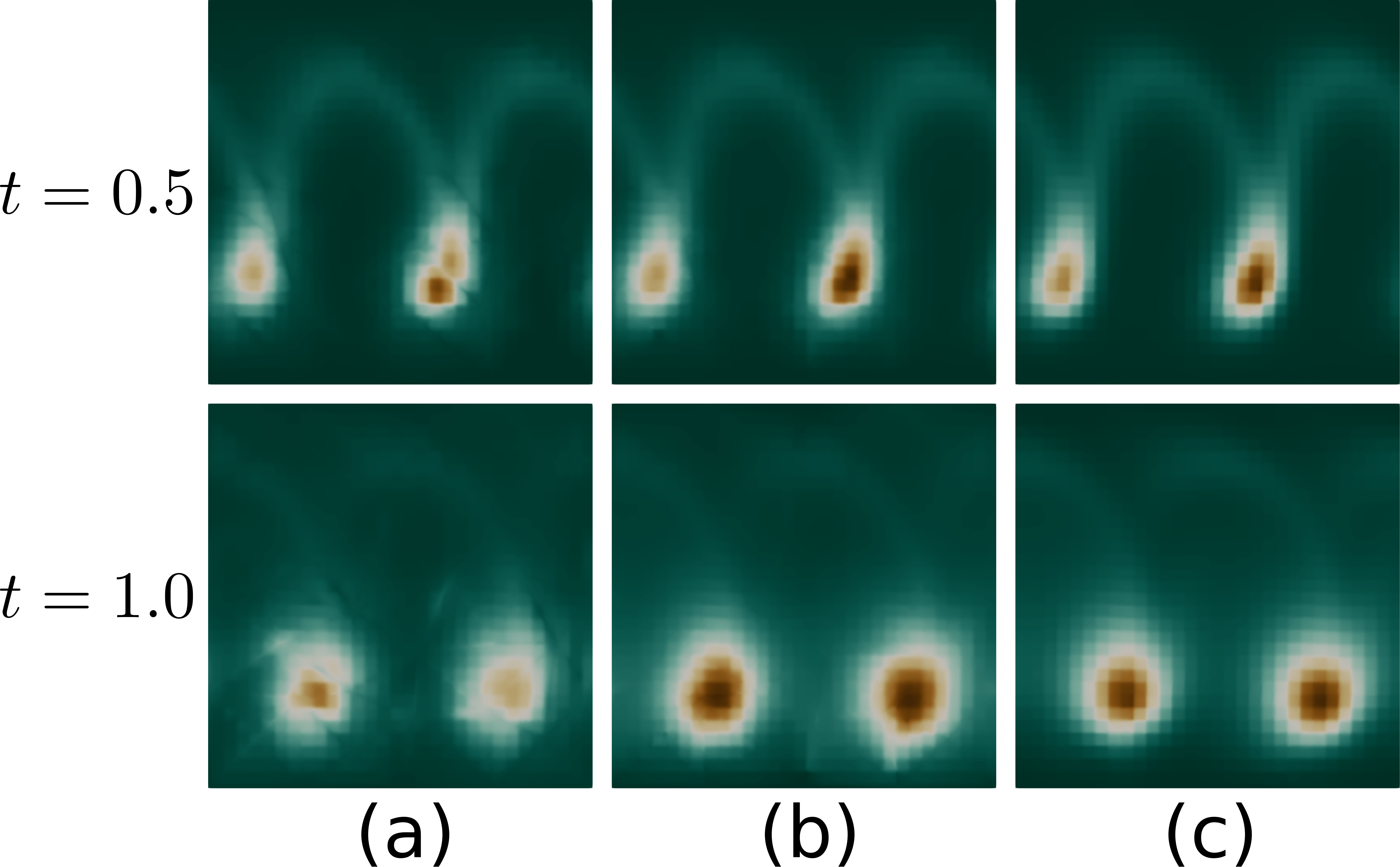}
  \end{center}
  \vspace{0pt}
  \caption{Comparison of \textbf{(a)} SLMsR without edge evolution and
    \textbf{(b)} SLMsR with edge evolution to \textbf{(c)} the
    reference solution in case of conservative problem \emph{Test 2b}
    shown at two time instances. The SLMsR without edge evolution
    shows stronger grid imprinting of the coarse mesh.}
  \vspace{0pt}
  \label{fig-2d_GaussDivConserv}
\end{figure}

Note that the difficulty we focus on here is not the actual
conservation (although this is an important issue) but the fact that
we do have another lower order (reaction) term in the equation with a
multiscale character. Fortunately this term is very local and is
handled nicely by the SLMsR since it scales with its magnitude. The
results are shown in Figure~\ref{fig-2d_GaussDivConserv} where we also
show how the quality of the SLMsR solution is negatively affected when
the reconstructed basis boundary conditions are not propagated. As the
reader can see the SLMsR without edge evolution shows a slight grid
imprinting of the coarse mesh. This effect can be more pronounced
depending on the strength of the reactive term since it is due to
keeping the boundary conditions fixed in time in the local basis
evolution step. This in turn increases the error in the evo-

\clearpage
\clearpage

\noindent lution of the asymptotic structure near coarse grid
boundaries due to the additional reaction term.

We pass on showing a comparison of all snapshots to the low resolution
FEM here (which does not behave well) for the sake of brevity. Error
plots including the low resolution FEM are again summarized in
Figure~\ref{fig-2d_error}.

\begin{figure}[t!]
  \vspace{-0pt}
  \begin{center}
    \includegraphics[width=1.0\textwidth]{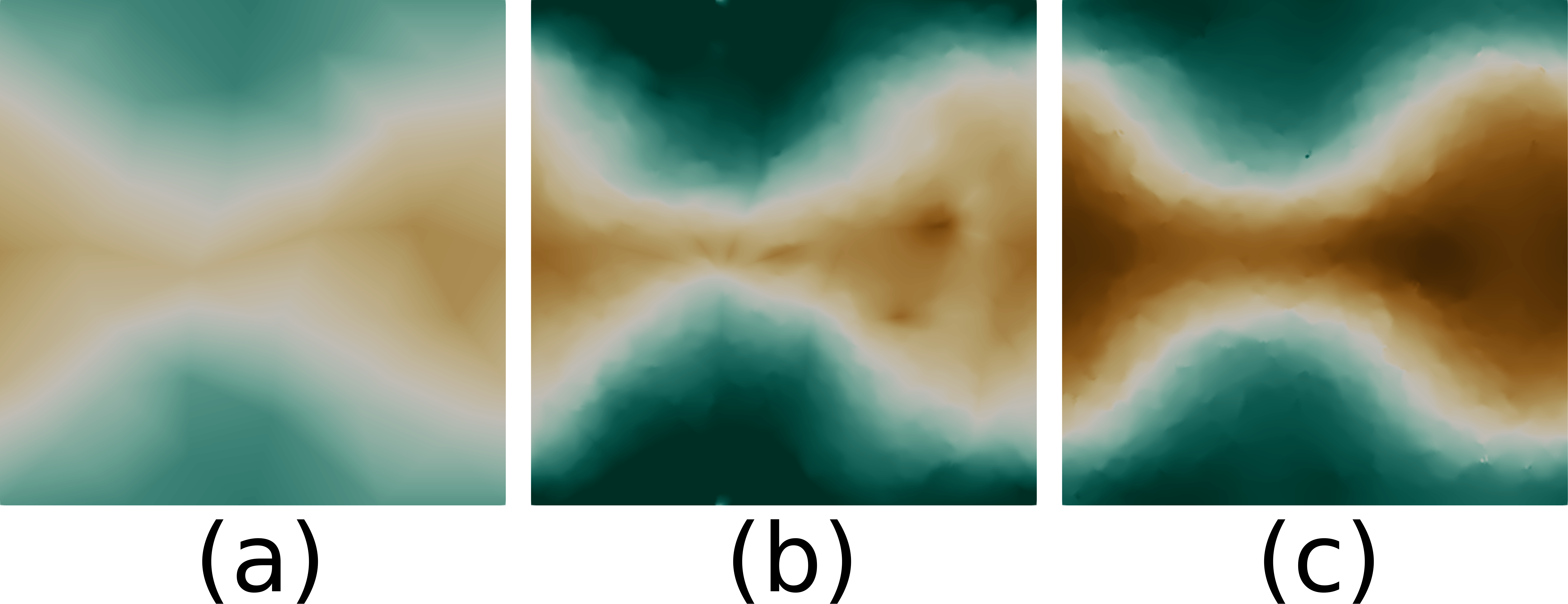}
  \end{center}
  \vspace{0pt}
  \caption{Snapshot of solutions of \emph{Test 3} at $T=1$. We show
    \textbf{(a)} the low resolution FEM, \textbf{(b)} the SLMsR and
    \textbf{(c)} the reference solution.}
  \vspace{0pt}
  \label{fig-2d_GaussRandom}
\end{figure}

\begin{wrapfigure}[18]{r}{0.6\textwidth}
  \vspace{0pt}
  \begin{center}
    \includegraphics[width=0.6\textwidth]{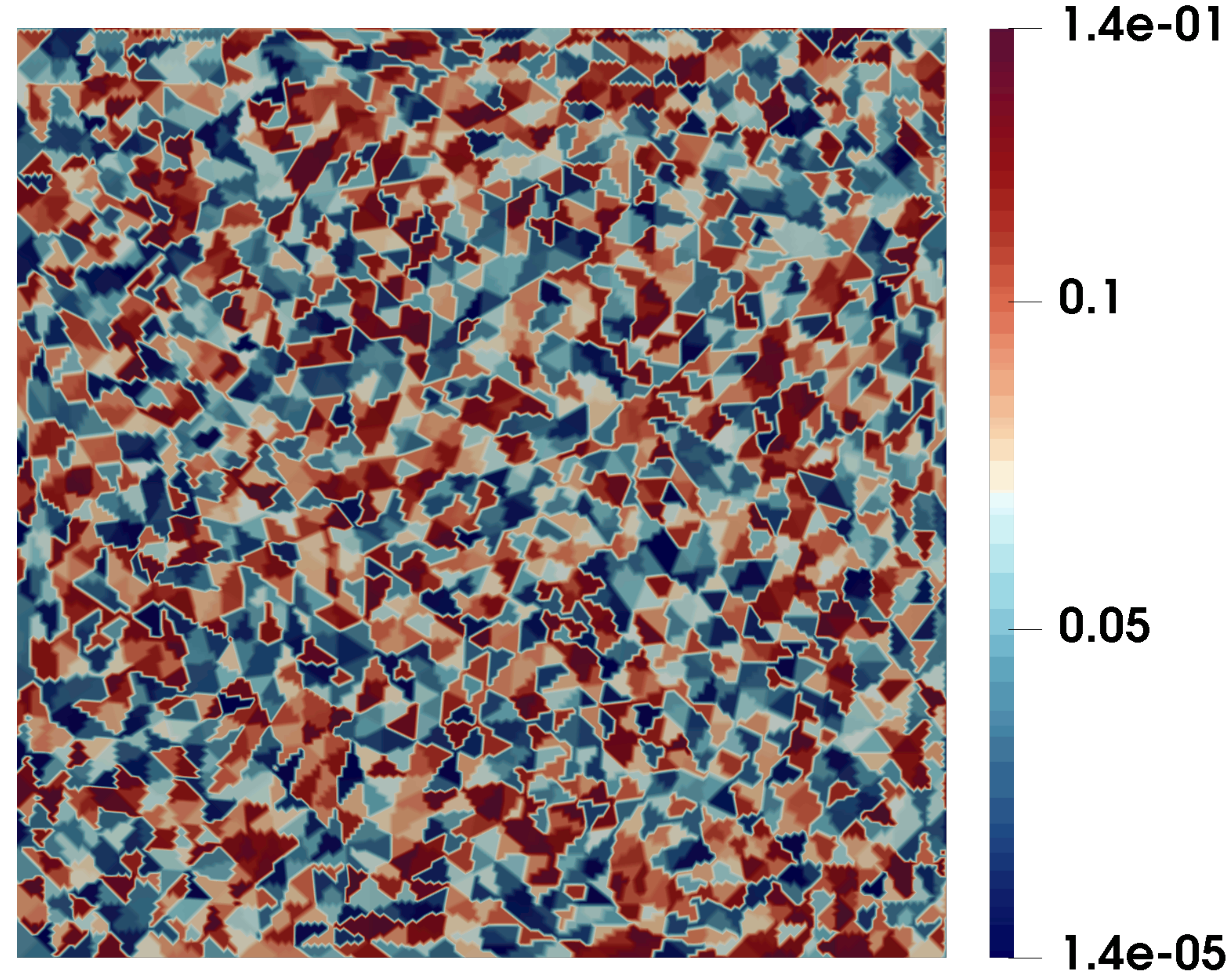}
  \end{center}
  \vspace{0pt}
  \caption{Magnitude of randomly generated (but fixed) diffusion
    tensor for \emph{Test 3}.}
  \vspace{0pt}
  \label{fig-2d_GaussRandomDiff}
\end{wrapfigure}

\paragraph{Test 3} This test has a similar intention as \emph{Test 3}
in Section~\ref{s-3}. We intend to show a reasonable behavior of the
SLMsR if the coefficients do not show a clear scale separation on
equation~(\ref{eq:2-1a}). For this we again used~(\ref{eq:4-4}) as
background velocity and generated a diagonal diffusion tensor that is
constant in each cell of a mesh generated to represent solely the
diffusion. The cell based constants are random and are fixed at the
beginning of the simultion. The diffusion tensor was then scaled to
contain values in the range of $10^{-5}$ to $10^{-1}$, see
Figure~\ref{fig-2d_GaussRandomDiff}. Snapshots of the solution can be
found in Figure~\ref{fig-2d_GaussRandom} and error plots in
Figure~\ref{fig-2d_error}.

\section{Summary and Discussion}\label{s-5}

In this work we introduced a multiscale method for advection-diffusion
equations that are dominated by the advective term. Such methods are
of importance, for example, in reservoir modeling and tracer transport
in earth system models. The main obstacles in these applications are,
first,  the advection-dominance and,
secondly, the multiscale character of the background velocity and the
diffusion tensor. The latter makes it impossible to simulate with
standard methods due to computational constraints while simulating
using standard methods with lower resolution that does not resolve
variations in the coefficients leads to wrong effective solutions.

Our idea to cope with these difficulties is inspired by ideas for
semi-Lagrangian methods, ideas based on ``convected fluid
microstucture'' as described in~\cite{Holm2012}, inverse problems and
multiscale finite elements~\cite{Efendiev2009}. At each time step we
reconstruct fine scale information from the solution at the previous
time step. This fine scale information enters the local representation
of the solution in each coarse cell, i.e., it is added as a corrector
to the local basis such that the basis representation is optimal in
some sense. Due to advection-dominance the optimal basis needs to be
reconstructed in its departure area, i.e., we locally trace back
information like in a semi-Lagrangian method for the
reconstruction. The reconstruction is performed by solving an inverse
problem with a suitable regularizer. It constructs a basis that does
not constitute a partition of unity (PoU) and is tailored for the
actual problem at hand. The idea of adding prior knowledge about the
solution to a local representation in PoU methods, however, is
similar, see for example~\cite{Melenk1996,Henning2014}. After
reconstructing the basis at the previous time step it is evolved with
suitable boundary conditions to the time level, where it is needed;
i.e., we evolve the local representation of the solution rather that
the solution itself. Note that the global framework of the SLMsR is
completely Eulerian while only the local reconstruction step in each
coarse cell is semi-Lagrangian.

The SLMsR also shares shallow similarities with stabilized methods
such as streamline upwind Petrov-Galerkin (SUPG) methods in which
``advected'' basis functions are used as correctors for test functions
to add a small diffusion in the direction of the flow. This is however
not the same since our basis also corrects for the fine scale
asymptotic structure induced by diffusion and other low order terms.

Numerical experiments that we carried out in one and two dimensions for
conservative and non-conservative advection diffusion equations show a
clear advantage in terms of accuracy for the SLMsR at low (coarse)
resolutions. We formulated the method globally as a conformal FEM to
show how the idea works in a quite restrictive setting. Other
frameworks are possible. A non-conformal FEM, for example, would put
less restrictions on the basis reconstruction at the price of having
to compute a boundary integral in the global assembly for each cell.

One of the main features of the SLMsR is its scalability: Although it
sounds expensive to trace back each coarse cell, then solve an inverse
problem and then solve a PDE at each time step (the so-called offline
phase) we would like to point out that these local problems are
independent and usually small and therefore the offline phase is
embaressingly parallel, although we did not take advantage of that in
our implementation yet. The global time step (online phase) also
consists of a small problem and matrix assembly procedures can be made
very efficient by using algebraic expressions,
see~\cite{HouWu1997,Efendiev2009}.

We would like to further emphasize the flexibility of the SLMsR. Here
we presented an implicit version but explicit time stepping is
possible. Also, the method can be transferred to higher dimensions as
well as it can be extended to deal with advection-diffusion-reaction
problems. Furthermore, the use of inverse problems in the local steps
to adjust the basis makes it generally possible to incorporate
knowledge coming from measurement data. For this a thorough
understanding of the data is necessary (as for any other assimilation
method). We would like to explore that opportunity in the future.

We believe that this method has potential to improve solutions to
real-life multi-scale problems since it is flexible and rises many
possibilities to transfer ideas to different types of equations and
methods. Some of the future issues that we aim at include its transfer
to other Galerkin methods such as discontinuous Galerkin methods,
attacking nonlinear and vector valued problems and at the use of
actual data for subscale assimilation.

\bibliography{reference}{}

\newpage
\appendix
\section{Tables}

  \begin{table}[h!]
    \begin{center}
      \begin{tabular}{ 
        p{0.05\linewidth} p{0.05\linewidth} | 
        p{0.075\textwidth} | p{0.075\textwidth} || 
        p{0.075\textwidth} | p{0.075\textwidth} }
        
        & & \multicolumn{2}{c||}{$L_{\mathrm{rel}}^2$}
        & \multicolumn{2}{c||}{$H_{\mathrm{rel}}^1$} \\
        
        \multicolumn{1}{c}{$H$} & \multicolumn{1}{c|}{$t$}
        & \multicolumn{1}{c|}{FEM} & \multicolumn{1}{c||}{SLMsR}
        & \multicolumn{1}{c|}{FEM} & \multicolumn{1}{c||}{SLMsR} \\ \hline
        
        \multicolumn{1}{c}{$H=1/8$}  & \multicolumn{1}{c|}{$1/3$} 
        & \multicolumn{1}{c|}{$0.109517$} & \multicolumn{1}{c||}{$0.0170849$} 
        & \multicolumn{1}{c|}{$0.738011$} & \multicolumn{1}{c||}{$0.223987$} \\

        & \multicolumn{1}{c|}{$2/3$} 
        & \multicolumn{1}{c|}{$0.110731$} & \multicolumn{1}{c||}{$0.0121468$} 
        & \multicolumn{1}{c|}{$0.521164$} & \multicolumn{1}{c||}{$0.104679$} \\

        & \multicolumn{1}{c|}{$1$} 
        & \multicolumn{1}{c|}{$0.100034$} & \multicolumn{1}{c||}{$0.0183614$} 
        & \multicolumn{1}{c|}{$0.652919$} & \multicolumn{1}{c||}{$0.191475$} \\ \hline

        \multicolumn{1}{c}{$H=1/32$}  & \multicolumn{1}{c|}{$1/3$} 
        & \multicolumn{1}{c|}{$0.0429804$} & \multicolumn{1}{c||}{$0.0141442$} 
        & \multicolumn{1}{c|}{$0.681634$} & \multicolumn{1}{c||}{$0.305783$} \\

        & \multicolumn{1}{c|}{$2/3$} 
        & \multicolumn{1}{c|}{$0.0336967$} & \multicolumn{1}{c||}{$0.00553025$}
        & \multicolumn{1}{c|}{$0.27492$} & \multicolumn{1}{c||}{$0.139942$} \\

        & \multicolumn{1}{c|}{$1$} 
        & \multicolumn{1}{c|}{$0.0264641$} & \multicolumn{1}{c||}{$0.00911659$}
        & \multicolumn{1}{c|}{$0.559342$} & \multicolumn{1}{c||}{$0.144821$}  \\ \hline

        \multicolumn{1}{c}{$H=1/128$}  & \multicolumn{1}{c|}{$1/3$} 
        & \multicolumn{1}{c|}{$0.0148088$} & \multicolumn{1}{c||}{$0.00477275$}
        & \multicolumn{1}{c|}{$0.431396$} & \multicolumn{1}{c||}{$0.17528$} \\ 
        
        & \multicolumn{1}{c|}{$2/3$} 
        & \multicolumn{1}{c|}{$0.0163083$} & \multicolumn{1}{c||}{$0.00225746$}
        & \multicolumn{1}{c|}{$0.173665$} & \multicolumn{1}{c||}{$0.0740701$}\\ 
        
        & \multicolumn{1}{c|}{$1$} 
        & \multicolumn{1}{c|}{$0.0113428$} & \multicolumn{1}{c||}{$0.00258289$}
        & \multicolumn{1}{c|}{$0.426786$} & \multicolumn{1}{c||}{$0.160761$} \\ \hline

        \multicolumn{1}{c}{$H=1/512$}  & \multicolumn{1}{c|}{$1/3$} 
        & \multicolumn{1}{c|}{$0.00088987$} & \multicolumn{1}{c||}{$0.00097043$} 
        & \multicolumn{1}{c|}{$0.126697$} & \multicolumn{1}{c||}{$0.0947134$} \\ 
        
        & \multicolumn{1}{c|}{$2/3$} 
        & \multicolumn{1}{c|}{$0.000589497$} & \multicolumn{1}{c||}{$0.000504804$} 
        & \multicolumn{1}{c|}{$0.0446262$} & \multicolumn{1}{c||}{$0.0435349$} \\ 
        
        & \multicolumn{1}{c|}{$1$} 
        & \multicolumn{1}{c|}{$0.000578547$} & \multicolumn{1}{c||}{$0.000491731$} 
        & \multicolumn{1}{c|}{$0.159977$} & \multicolumn{1}{c||}{$0.173261$} \\ \hline

        \hline
      \end{tabular}
    
      \vspace{1cm}

      \begin{tabular}{ 
        p{0.05\linewidth} p{0.05\linewidth} | 
        p{0.075\textwidth} | p{0.075\textwidth} || 
        p{0.075\textwidth} | p{0.075\textwidth} }
        
        & & \multicolumn{2}{c||}{$L_{\mathrm{rel}}^2$}
        & \multicolumn{2}{c||}{$H_{\mathrm{rel}}^1$} \\
        
        \multicolumn{1}{c}{$H$} & \multicolumn{1}{c|}{$t$}
        & \multicolumn{1}{c|}{FEM} & \multicolumn{1}{c||}{SLMsR}
        & \multicolumn{1}{c|}{FEM} & \multicolumn{1}{c|}{SLMsR} \\ \hline
        
        \multicolumn{1}{c}{$H=1/8$}  & \multicolumn{1}{c|}{$1/3$} 
        & \multicolumn{1}{c|}{$0.214453$} & \multicolumn{1}{c||}{$0.0702007$} 
        & \multicolumn{1}{c|}{$0.921891$} & \multicolumn{1}{c||}{$0.321898$} \\ 

        & \multicolumn{1}{c|}{$2/3$} 
        & \multicolumn{1}{c|}{$0.171949$} & \multicolumn{1}{c||}{$0.0662604$} 
        & \multicolumn{1}{c|}{$0.940582$} & \multicolumn{1}{c||}{$0.261431$} \\ 

        & \multicolumn{1}{c|}{$1$} 
        & \multicolumn{1}{c|}{$0.188325$} & \multicolumn{1}{c||}{$0.0743565$} 
        & \multicolumn{1}{c|}{$0.975117$} & \multicolumn{1}{c||}{$0.336754$} \\ \hline

        \multicolumn{1}{c}{$H=1/32$}  & \multicolumn{1}{c|}{$1/3$} 
        & \multicolumn{1}{c|}{$0.139444$} & \multicolumn{1}{c||}{$0.0283479$} 
        & \multicolumn{1}{c|}{$0.975416$} & \multicolumn{1}{c||}{$0.256681$} \\

        & \multicolumn{1}{c|}{$2/3$} 
        & \multicolumn{1}{c|}{$0.108827$} & \multicolumn{1}{c||}{$0.0294402$}
        & \multicolumn{1}{c|}{$0.974576$} & \multicolumn{1}{c||}{$0.250716$} \\

        & \multicolumn{1}{c|}{$1$} 
        & \multicolumn{1}{c|}{$0.140284$} & \multicolumn{1}{c||}{$0.0398691$}
        & \multicolumn{1}{c|}{$1.02447$} & \multicolumn{1}{c||}{$0.252878$} \\ \hline

        \multicolumn{1}{c}{$H=1/128$}  & \multicolumn{1}{c|}{$1/3$} 
        & \multicolumn{1}{c|}{$0.0373321$} & \multicolumn{1}{c||}{$0.00634348$}
        & \multicolumn{1}{c|}{$0.62717$} & \multicolumn{1}{c||}{$0.109299$} \\ 
        
        & \multicolumn{1}{c|}{$2/3$} 
        & \multicolumn{1}{c|}{$0.0322432$} & \multicolumn{1}{c||}{$0.0119677$}
        & \multicolumn{1}{c|}{$0.628683$} & \multicolumn{1}{c||}{$0.086647$} \\ 
        
        & \multicolumn{1}{c|}{$1$} 
        & \multicolumn{1}{c|}{$0.0446856$} & \multicolumn{1}{c||}{$0.0233804$}
        & \multicolumn{1}{c|}{$0.78208$} & \multicolumn{1}{c||}{$0.183882$} \\ \hline

        \multicolumn{1}{c}{$H=1/512$}  & \multicolumn{1}{c|}{$1/3$} 
        & \multicolumn{1}{c|}{$0.00274732$} & \multicolumn{1}{c||}{$0.00200674$} 
        & \multicolumn{1}{c|}{$0.182691$} & \multicolumn{1}{c||}{$0.0595046$} \\ 
        
        & \multicolumn{1}{c|}{$2/3$} 
        & \multicolumn{1}{c|}{$0.00238125$} & \multicolumn{1}{c||}{$0.00412889$} 
        & \multicolumn{1}{c|}{$0.181782$} & \multicolumn{1}{c||}{$0.0554508$} \\ 
        
        & \multicolumn{1}{c|}{$1$} 
        & \multicolumn{1}{c|}{$0.0045627$} & \multicolumn{1}{c||}{$0.00622009$} 
        & \multicolumn{1}{c|}{$0.290947$} & \multicolumn{1}{c||}{$0.140458$} \\ \hline

        \hline
      \end{tabular}
    \end{center}
    
    \caption{Relative errors of standard FEM and SLMsR 
      for the non-conservative test problem~(\ref{eq:2-16}) (upper table) 
      and for the conservative problem~(\ref{eq:2-17}) (lower table). The 
      errors are shown at three different time steps.}

    \label{tabl-1d_convergence_test}
  \end{table}

  \begin{table}[h!]
    \begin{center}
      \begin{tabular}{ 
        p{0.05\linewidth} p{0.05\linewidth} | 
        p{0.075\textwidth} | p{0.075\textwidth} || 
        p{0.075\textwidth} | p{0.075\textwidth} }
        
        & & \multicolumn{2}{c||}{$L_{\mathrm{rel}}^2$}
        & \multicolumn{2}{c||}{$H_{\mathrm{rel}}^1$} \\
        
        \multicolumn{1}{c}{$H$} & \multicolumn{1}{c|}{$t$}
        & \multicolumn{1}{c|}{FEM} & \multicolumn{1}{c||}{SLMsR}
        & \multicolumn{1}{c|}{FEM} & \multicolumn{1}{c||}{SLMsR} \\ \hline
        
        \multicolumn{1}{c}{$  1/8$}  & \multicolumn{1}{c|}{$1/2$} 
        & \multicolumn{1}{c|}{$2.71124\E{-1}$} & \multicolumn{1}{c||}{$3.29375\E{-2}$} 
        & \multicolumn{1}{c|}{$9.47533\E{-1}$} & \multicolumn{1}{c||}{$4.26613\E{-1}$} \\

        & \multicolumn{1}{c|}{$1$} 
        & \multicolumn{1}{c|}{$4.80502\E{-1}$} & \multicolumn{1}{c||}{$1.9577\E{-2}$} 
        & \multicolumn{1}{c|}{$1.07721\E{0}$} & \multicolumn{1}{c||}{$3.62467\E{-1}$} \\ \hline

        \multicolumn{1}{c}{$1/32$}  & \multicolumn{1}{c|}{$1/2$}
        & \multicolumn{1}{c|}{$2.38448\E{-1}$} & \multicolumn{1}{c||}{$1.51868\E{-2}$} 
        & \multicolumn{1}{c|}{$8.06106\E{-1}$} & \multicolumn{1}{c||}{$2.71302\E{-1}$} \\

        & \multicolumn{1}{c|}{$1$} 
        & \multicolumn{1}{c|}{$3.4961\E{-1}$} & \multicolumn{1}{c||}{$2.03215\E{-2}$}
        & \multicolumn{1}{c|}{$1.038\E{0}$} & \multicolumn{1}{c||}{$2.62539\E{-1}$}  \\ \hline

        \multicolumn{1}{c}{$1/128$}  & \multicolumn{1}{c|}{$1/2$} 
        & \multicolumn{1}{c|}{$8.23856\E{-2}$} & \multicolumn{1}{c||}{$1.07117\E{-2}$}
        & \multicolumn{1}{c|}{$6.75349\E{-1}$} & \multicolumn{1}{c||}{$2.62084\E{-1}$} \\ 
        
        & \multicolumn{1}{c|}{$1$} 
        & \multicolumn{1}{c|}{$1.52502\E{-1}$} & \multicolumn{1}{c||}{$2.39788\E{-2}$}
        & \multicolumn{1}{c|}{$6.54747\E{-1}$} & \multicolumn{1}{c||}{$2.84093\E{-1}$} \\ \hline

        \multicolumn{1}{c}{$1/512$}  & \multicolumn{1}{c|}{$1/2$} 
        & \multicolumn{1}{c|}{$1.48402\E{-2}$} & \multicolumn{1}{c||}{$1.74951\E{-3}$} 
        & \multicolumn{1}{c|}{$3.35263\E{-1}$} & \multicolumn{1}{c||}{$2.24681\E{-1}$} \\ 
        
        & \multicolumn{1}{c|}{$1$} 
        & \multicolumn{1}{c|}{$2.85318\E{-2}$} & \multicolumn{1}{c||}{$1.65701\E{-3}$} 
        & \multicolumn{1}{c|}{$3.33875\E{-1}$} & \multicolumn{1}{c||}{$2.54352\E{-1}$} \\

        \hline
      \end{tabular}

      \vspace{0.5cm}
      \begin{tabular}{ 
        p{0.05\linewidth} p{0.05\linewidth} | 
        p{0.075\textwidth} | p{0.075\textwidth} || 
        p{0.075\textwidth} | p{0.075\textwidth} }
        
        & & \multicolumn{2}{c||}{EOC $L_{\mathrm{rel}}^2$}
        & \multicolumn{2}{c||}{EOC $H_{\mathrm{rel}}^1$} \\
        
        \multicolumn{1}{c}{$H\rightarrow H/4$} & \multicolumn{1}{c|}{$h\rightarrow h/4$}
        & \multicolumn{1}{c|}{FEM} & \multicolumn{1}{c||}{SLMsR}
        & \multicolumn{1}{c|}{FEM} & \multicolumn{1}{c||}{SLMsR} \\ \hline
        
        \multicolumn{1}{c}{$2^{-3}\rightarrow 2^{-5}$}  & \multicolumn{1}{c|}{$2^{-9}\rightarrow 2^{-11}$} 
        & \multicolumn{1}{c|}{$0.22939$} & \multicolumn{1}{c||}{$-0.02692$} 
        & \multicolumn{1}{c|}{$0.02674$} & \multicolumn{1}{c||}{$0.23265$} \\

        \multicolumn{1}{c}{$2^{-5}\rightarrow 2^{-7}$}  & \multicolumn{1}{c|}{$2^{-11}\rightarrow 2^{-13}$} 
        & \multicolumn{1}{c|}{$0.59845$} & \multicolumn{1}{c||}{$-0.11937$} 
        & \multicolumn{1}{c|}{$0.33239$} & \multicolumn{1}{c||}{$-0.05691$} \\

        \multicolumn{1}{c}{$2^{-7}\rightarrow 2^{-9}$}  & \multicolumn{1}{c|}{$2^{-13}\rightarrow 2^{-15}$} 
        & \multicolumn{1}{c|}{$1.20909$} & \multicolumn{1}{c||}{$1.92755$} 
        & \multicolumn{1}{c|}{$0.48581$} & \multicolumn{1}{c||}{$0.07977$} \\

        \hline
      \end{tabular}
    \end{center}
    
    \caption{\textbf{Upper table:} relative errors at time $t=1/2$ and $t=1$ for the unresolved 
      regime of test problem~(\ref{eq:2-19}) in $L^2(\tor^1)$ and $H^1(\tor^1)$
      at different coarse resolutions. The fine resolution was fixed to $n_f=64$ cells 
      per coarse cell. \textbf{Lower table:} estimated order of convegence at $t=1$.}

    \label{tabl-1d_convergence_unresolved}
  \end{table}
  \begin{table}[h!]
    \begin{center}
      \begin{tabular}{ 
        p{0.05\linewidth} p{0.05\linewidth} | 
        p{0.075\textwidth} | p{0.075\textwidth} || 
        p{0.075\textwidth} | p{0.075\textwidth} }
        
        & & \multicolumn{2}{c||}{$L_{\mathrm{rel}}^2$}
        & \multicolumn{2}{c||}{$H_{\mathrm{rel}}^1$} \\
        
        \multicolumn{1}{c}{$H$} & \multicolumn{1}{c|}{$t$}
        & \multicolumn{1}{c|}{FEM} & \multicolumn{1}{c||}{SLMsR}
        & \multicolumn{1}{c|}{FEM} & \multicolumn{1}{c||}{SLMsR} \\ \hline
        
        \multicolumn{1}{c}{$  1/16$}  & \multicolumn{1}{c|}{$1/2$} 
        & \multicolumn{1}{c|}{$3.14117\E{-2}$} & \multicolumn{1}{c||}{$1.83944\E{-3}$} 
        & \multicolumn{1}{c|}{$2.16986\E{-1}$} & \multicolumn{1}{c||}{$2.17566\E{-2}$} \\

        & \multicolumn{1}{c|}{$1$} 
        & \multicolumn{1}{c|}{$2.71271\E{-2}$} & \multicolumn{1}{c||}{$2.16896\E{-3}$} 
        & \multicolumn{1}{c|}{$1.62237\E{-1}$} & \multicolumn{1}{c||}{$1.58667\E{-2}$} \\ \hline

        \multicolumn{1}{c}{$1/32$}  & \multicolumn{1}{c|}{$1/2$}
        & \multicolumn{1}{c|}{$7.61179\E{-3}$} & \multicolumn{1}{c||}{$6.45978\E{-4}$} 
        & \multicolumn{1}{c|}{$1.064\E{-1}$} & \multicolumn{1}{c||}{$2.4219\E{-2} $} \\

        & \multicolumn{1}{c|}{$1$} 
        & \multicolumn{1}{c|}{$6.31955\E{-3}$} & \multicolumn{1}{c||}{$5.84086\E{-4}$}
        & \multicolumn{1}{c|}{$7.45369\E{-2}$} & \multicolumn{1}{c||}{$1.58233\E{-2}$}  \\ \hline

        \multicolumn{1}{c}{$1/64$}  & \multicolumn{1}{c|}{$1/2$} 
        & \multicolumn{1}{c|}{$1.90004\E{-3}$} & \multicolumn{1}{c||}{$3.36293\E{-4}$}
        & \multicolumn{1}{c|}{$5.32668\E{-2}$} & \multicolumn{1}{c||}{$2.61811\E{-2}$} \\ 
        
        & \multicolumn{1}{c|}{$1$} 
        & \multicolumn{1}{c|}{$1.56285\E{-3}$} & \multicolumn{1}{c||}{$2.41687\E{-4}$}
        & \multicolumn{1}{c|}{$3.68329\E{-2}$} & \multicolumn{1}{c||}{$1.72855\E{-2}$} \\ \hline

        \multicolumn{1}{c}{$1/128$}  & \multicolumn{1}{c|}{$1/2$} 
        & \multicolumn{1}{c|}{$4.73831\E{-4}$} & \multicolumn{1}{c||}{$1.51248\E{-4}$} 
        & \multicolumn{1}{c|}{$2.66104\E{-2}$} & \multicolumn{1}{c||}{$2.2229\E{-2}$} \\ 
        
        & \multicolumn{1}{c|}{$1$} 
        & \multicolumn{1}{c|}{$3.88823\E{-4}$} & \multicolumn{1}{c||}{$1.07755\E{-4}$} 
        & \multicolumn{1}{c|}{$1.83422\E{-2}$} & \multicolumn{1}{c||}{$1.55475\E{-2}$} \\ \hline

        \multicolumn{1}{c}{$1/256$}  & \multicolumn{1}{c|}{$1/2$} 
        & \multicolumn{1}{c|}{$1.17218\E{-4}$} & \multicolumn{1}{c||}{$4.54382\E{-5}$} 
        & \multicolumn{1}{c|}{$1.32288\E{-2}$} & \multicolumn{1}{c||}{$ 1.31489\E{-2}$} \\ 
        
        & \multicolumn{1}{c|}{$1$} 
        & \multicolumn{1}{c|}{$9.60583\E{-5}$} & \multicolumn{1}{c||}{$3.19474\E{-5}$} 
        & \multicolumn{1}{c|}{$9.11126\E{-3}$} & \multicolumn{1}{c||}{$9.05413\E{-3}$} \\

        \hline
      \end{tabular}

      \vspace{0.5cm}
      \begin{tabular}{ 
        p{0.05\linewidth} p{0.05\linewidth} | 
        p{0.075\textwidth} | p{0.075\textwidth} || 
        p{0.075\textwidth} | p{0.075\textwidth} }
        
        &  & \multicolumn{2}{c||}{EOC $L_{\mathrm{rel}}^2$}
        & \multicolumn{2}{c||}{EOC $H_{\mathrm{rel}}^1$} \\
        
        \multicolumn{1}{c}{$H\rightarrow H/4$} & \multicolumn{1}{c|}{$h\rightarrow h/4$}
        & \multicolumn{1}{c|}{FEM} & \multicolumn{1}{c||}{SLMsR}
        & \multicolumn{1}{c|}{FEM} & \multicolumn{1}{c||}{SLMsR} \\ \hline
        
        \multicolumn{1}{c}{$2^{-4}\rightarrow 2^{-5}$}  & \multicolumn{1}{c|}{$2^{-9}\rightarrow 2^{-10}$} 
        
        & \multicolumn{1}{c|}{$2.10184$} & \multicolumn{1}{c||}{$1.89275$} 
        & \multicolumn{1}{c|}{$1.12208$} & \multicolumn{1}{c||}{$0.003951$} \\ 

        \multicolumn{1}{c}{$2^{-5}\rightarrow 2^{-6}$}  & \multicolumn{1}{c|}{$2^{-10}\rightarrow 2^{-11}$} 
        & \multicolumn{1}{c|}{$2.01564$} & \multicolumn{1}{c||}{$ 1.27304$} 
        & \multicolumn{1}{c|}{$1.01696$} & \multicolumn{1}{c||}{$-0.12751$} \\ 

        \multicolumn{1}{c}{$2^{-6}\rightarrow 2^{-7}$}  & \multicolumn{1}{c|}{$2^{-11}\rightarrow 2^{-12}$} 
        & \multicolumn{1}{c|}{$2.00699$} & \multicolumn{1}{c||}{$1.16538$} 
        & \multicolumn{1}{c|}{$1.00583$} & \multicolumn{1}{c||}{$0.152887$} \\

        \multicolumn{1}{c}{$2^{-7}\rightarrow 2^{-8}$}  & \multicolumn{1}{c|}{$2^{-12}\rightarrow 2^{-13}$} 
        & \multicolumn{1}{c|}{$2.01713$} & \multicolumn{1}{c||}{$1.75399$} 
        & \multicolumn{1}{c|}{$1.00944$} & \multicolumn{1}{c||}{$0.780031$} \\

        \multicolumn{1}{c}{$2^{-8}\rightarrow 2^{-9}$}  & \multicolumn{1}{c|}{$2^{-13}\rightarrow 2^{-14}$} 
        & \multicolumn{1}{c|}{$2.06791$} & \multicolumn{1}{c||}{$2.13616$} 
        & \multicolumn{1}{c|}{$1.03542$} & \multicolumn{1}{c||}{$1.02688$} \\ 

        \hline
      \end{tabular}
    \end{center}
    \caption{\textbf{Upper table:} relative errors at time $t=1/2$ and $t=1$ for the resolved 
      regime of test problem~(\ref{eq:2-20}) in $L^2(\tor^1)$ and $H^1(\tor^1)$
      at different coarse resolutions. The fine resolution was fixed to $n_f=32$ cells 
      per coarse cell. \textbf{lower table:} estimated order of convergence at $t=1$.}
    \label{tabl-1d_convergence_resolved}
  \end{table}

\end{document}